\documentclass[11pt]{article}

\usepackage{amsmath,amsfonts,graphicx,epsfig,mathrsfs,amssymb,tikz,float}
\usepackage{bm}
\usepackage{bbm}
\usepackage{dsfont}
\usepackage{mathtools,mathbbol}
\usepackage{color}
\usepackage{tensor}
\usepackage[colorlinks]{hyperref}
\hypersetup{linktocpage}
\usepackage[titletoc]{appendix}
\usepackage{cite}
\usepackage{ragged2e}
\usepackage{upgreek}
\usepackage{physics}
\usepackage[normalem]{ulem}
\usepackage{url}
\usepackage{setspace}
\usepackage[font=footnotesize]{caption}
\usepackage{textcomp}
\usepackage{subcaption}
\usepackage{graphicx}
\pdfoutput=1

\definecolor{carmine}{rgb}{0.59, 0.0, 0.09}
\definecolor{sgreen}{rgb}{0.0, 0.44, 0.0}
\definecolor{prussianblue}{rgb}{0.0, 0.06, 0.54}
\hypersetup{colorlinks=true, linkcolor=carmine, filecolor=magenta, urlcolor=prussianblue, citecolor=sgreen}

\def\bea{\begin{eqnarray}}
\def\eea{\end{eqnarray}}
\def\be{\begin{equation}}
\def\ee{\end{equation}}
\def\ba{\begin{array}}
\def\ea{\end{array}}

\newcommand{\x}{{\boldsymbol x}}
\newcommand{\y}{{\boldsymbol y}}
\newcommand{\K}{{\boldsymbol k}}
\newcommand{\q}{{\boldsymbol q}}

\newcommand{\de}{{\rm d}}

\def\nn{\nonumber}

\begin{document}

\setlength\arraycolsep{2pt}

\renewcommand{\theequation}{\arabic{section}.\arabic{equation}}
\setcounter{page}{1}

\begin{titlepage}

\begin{center}

\vskip 1.5 cm

{\huge\bf
Particle creation from non-geodesic trajectories in multifield inf{l}ation 
}

\vskip 2.0cm

{\large Nicol\'as Parra$^a$, Spyros Sypsas$^{b,c}$, Gonzalo A. Palma$^a$,  and Crist\'obal Zenteno$^d$}

\vskip .6cm

$^a$ {\it Departamento de Física, FCFM, Universidad de Chile,\\ Blanco Encalada 2008, Santiago, Chile}

$^b$  {\it High Energy Physics Research Unit, Faculty of Science, Chulalongkorn University, Bangkok 10330, Thailand}

$^c$  {\it National Astronomical Research Institute of Thailand, Don Kaeo, Mae Rim, Chiang Mai 50180, Thailand}

$^d$  {\it Instituto de Física Teórica UAM/CSIC, c/ Nicolás Cabrera 13-15, 28049, Madrid, Spain}

\vskip 2.5cm

\end{center}

\begin{abstract} 

Particle production in de Sitter spacetime arises from the exponential expansion of space, rendering the Bunch-Davies vacuum perceived as a particle-containing state by late-time observers. For states defined as eigenstates of both momentum and the Hamiltonian, the Bunch-Davies vacuum exhibits a constant particle density per physical momentum. We explore particle production beyond this baseline, focusing on deviations from exact de Sitter conditions and non-gravitational interactions, such as slow-roll inflation or interactions arising from the coupling of inflation to other fields. Using Bogoliubov transformations, we calculate the number density of energy/momentum eigenstates. In single-field inflation, this density captures the observed spectral index of the primordial power spectrum, while in two-field models, it reflects the non-gravitational coupling driving background trajectory turning. We present analytical results applicable to various scenarios involving particle production from non-adiabatic processes during inflation.

\end{abstract}

\end{titlepage}

\tableofcontents

\setcounter{equation}{0}
\section{Introduction}

In contrast to flat spacetime, where the vacuum state is well-defined and stable, curved spacetime introduces ambiguities in the definition of particles~\cite{Birrell:1982ix}, leading to situations where different observers may disagree on the particle content of a system~\cite{Parker:1968mv,Parker:1969au,Imamura:1960tzx,Parker:1971pt,Ford:2021syk,Kolb:2023ydq,Chung:1998zb,Chung:1998ua}. This effect is especially prominent in time-dependent spacetimes~\cite{Armendariz-Picon:2003knj}, such as cosmological backgrounds, where changes in the geometry can cause transitions in the system's quantum state, resulting in the spontaneous production of particles as perceived by a single observer at different times.

In the particular case of cosmic inflation~\cite{Starobinsky:1980te,Guth:1980zm,Linde:1981mu,Albrecht:1982wi}, particle production is usually understood as arising from the nearly exponential expansion of space, which causes the Bunch-Davies vacuum defined at early times to be perceived as a particle-containing state by late-time observers. For instance, for a massless scalar field in an exact de Sitter spacetime with spatially flat foliations and a constant expansion rate $H$, the Bunch-Davies (BD) state displays a phase-space number-density (expressed in terms of physical momenta $p$) that remains constant over time~\cite{Garbrecht:2002pd}:
\be
 n^{\rm BD}_{\boldsymbol{p}} =  \frac{H^2}{4 p^2} . \label{intro-n-BD}
\ee
This result gives the density of particles as $n = \int \dd[3] p \, n^{\rm BD}_{\boldsymbol{p}}$ and is computed for particle states defined to be eigenstates of the system's Hamiltonian.

This expression is valid for any momentum and is consistent with the fact that de Sitter spacetime does not favor any particular time. Consequently, particle production on top of this bath requires either departures from exact de Sitter conditions, such as those present in slow-roll inflation scenarios, or the presence of non-gravitational interactions, as typically encountered in multifield inflationary models. Of course, eventually inflation ends, leaving behind the nearly de Sitter stage characterized by (\ref{intro-n-BD}), and any quanta produced during inflation must have decayed into resultant products, heating up the universe. Counting with analytical expressions for the number-density of particles, such as (\ref{intro-n-BD}) or alterations thereof due to non-adiabatic dynamics, turns out to be relevant in order to infer properties associated to the end of inflation, even though a complete characterization is found to be somewhat model dependent. The purpose of this article is to analyze particle production during inflation paying special attention on small departures from~(\ref{intro-n-BD}) as a consequence of interactions found in multifield models of inflation.

A common approach to studying quantum particle production in time-dependent backgrounds, involves the calculation of Bogoliubov coefficients, typically denoted by $\alpha_k$ and $\beta_k$~\cite{Ford:2021syk,Kolb:2023ydq}. These coefficients parameterize a linear, canonical transformation known as a Bogoliubov transformation, which relates the vacuum state of the system before and after a background change. Equivalently, they describe how the field operator for a given particle species transforms in response to the changing background. More to the point, for a given wave-vector $\boldsymbol{k}$, the mode function $u^{\rm out}_k(t)$ after the background evolution can be expressed in terms of the ``positive frequency" mode function $u^{\rm in}_k(t)$ before the change takes place. This relationship is given by:
\be
u_k^{\text{out}}(t) = \alpha_k u_k^{\text{in}}(t) + \beta_k \left[ u_k^{\text{in}}(t) \right]^*, \label{basic-def-bogos}
\ee
where $\alpha_k$ represents the contribution from the original positive frequency mode, and $\beta_k$, which appears thanks to the background variation, quantifies the mixing with the negative frequency mode, corresponding to particle creation. For instance, in the case of a strong variation of the electromagnetic field, (\ref{basic-def-bogos}) gives the the phase-space number-density $n_{\boldsymbol{p}}$ of electron-positron pairs found to be proportional to $|\beta_k|^2$, a phenomenon known as the Schwinger effect; see~\cite{Martin:2007bw} for a direct comparison to the cosmological analogue.

In single-field slow-roll inflation, the background geometry is constantly changing over time due to the slowly varying inflaton field. As a consequence, the vacuum is permanently redefined, forcing one to consider time-dependent Bogoliubov coefficients. For instance, one possibility to study particle creation is to adapt (\ref{basic-def-bogos}) to the form
\be
u_k^{\text{out}}(t) = \alpha_k (t) \, u_k^{\text{BD}}(t) + \beta_k (t) \, \left[ u_k^{\text{BD}}(t) \right]^*, \label{basic-def-bogos-2}
\ee
where $u_k^{\text{BD}}(t)$ is the Bunch-Davies mode function describing fluctuations of the inflaton field or, working in comoving gauge, primordial curvature fluctuations. In this case, the wave-vector $\boldsymbol{k}$, which remains constant, is related to physical momentum $\boldsymbol{p}$ via ${\boldsymbol{p}} = {\boldsymbol{k}}/a$, where $a = a(t)$ is the scale factor describing how spatial slices expand at the Hubble rate $H = \dot a / a$. The Bunch-Davies mode function is chosen in such a way that for very early times it resembles a positive frequency mode in a Minkowski spacetime. In this description, the Bogoliubov coefficients are time-dependent, keeping track of continuous particle creation due to departures from the exact de Sitter geometry induced by the slowly changing expansion rate $H$. These departures are commonly parametrized by the dimensionless slow-roll parameters $\epsilon = - \dot H / H^2$ and $\eta = \dot \epsilon / H \epsilon$. It follows that $\alpha(t)$ and $\beta (t)$ collectively satisfy a first order differential equation of the form 
\be
\frac{\dd}{\dd t}\mqty( \alpha_k(t) \\ \beta_k(t) )  = \mathcal M_k (t) \mqty( \alpha_k(t) \\ \beta_k(t) ) , \label{intro-diff-alpha-beta}
\ee
where $\mathcal M_k (t)$ is a $2\times 2$ matrix proportional to $\epsilon$ and $\eta$. Similarly to the Schwinger effect, in this case the number density per momentum of particles produced on top of the de Sitter particle bath is $N_k(t) = |\beta_k (t)|^2$, which is found to be of order $\epsilon$ and $\eta$. 

As we shall see, within this scheme of studying particle production during inflation, the number-density of particles can be related to departures from scale invariance of the power spectrum. In single-field slow-roll inflation, on large cosmological scales (\emph{i.e.} scales that remained superhorizon during reheating), one finds that the number-density of particles produced in a quasi-de Sitter spacetime is
\be
 n^{\rm qdS}_{\boldsymbol{p}}   =    n^{\rm BD}_{\boldsymbol{p}} \times \left( \frac{p}{H} \right)^{n_s - 1}   ,  \label{density-single-field}
\ee
where $n^{\rm BD}_{\boldsymbol{p}}$ is the number-density of particles present in a rigid de Sitter spacetime, and $n_s - 1$ is the spectral index parametrizing the deviation of the power spectrum from scale invariance. 

From an effective field theory perspective, it is quite natural to expect a multitude of degrees of freedom to become operative at high energies. 
How does the previous discussion map into multifield inflation? The purpose of this article is to analyze particle creation in multifield models where the curvature fluctuation interacts with other isocurvature perturbations. In multifield models, the curvature fluctuation is associated to perturbations tangent to the trajectory followed by the inflaton's field vacuum expectation value in the multidimensional target space. On the contrary, isocurvature fluctuations correspond to perturbations normal to the trajectory, and may experience non-gravitational interactions with the curvature perturbation whenever the trajectory becomes non-geodesic (that is, when the trajectory experiences a turn in target space)~\cite{GrootNibbelink:2001qt,Achucarro:2010da,Cremonini:2010ua,Gong:2011uw,Renaux-Petel:2015mga}. As we shall see, in the simple case of two-field models, the time-dependent Bogoliubov coefficients follow a differential equation of the form
\be
\frac{\dd}{\dd t}\mqty( \alpha^{(1)}_k(t) \\ \beta^{(1)}_k(t) \\ \alpha^{(2)}_k(t) \\ \beta^{(2)}_k(t))  = \mathcal M_k (t) \mqty( \alpha^{(1)}_k(t) \\ \beta^{(1)}_k(t) \\ \alpha^{(2)}_k(t) \\ \beta^{(2)}_k(t)) ,
\ee
where $\alpha^{(i)}_k(t)$ and $\beta^{(i)}_k(t)$ are the Bogoliubov coefficients related to the $i$-th fluctuation. Disregarding slow-roll parameters, $\mathcal M_k (t)$ is now a $4\times 4$ matrix proportional to the rate of turn $\Omega$ of the non-geodesic multifield trajectory. This new set of Bogoliubov coefficients leads to several interesting and useful results, such as the number density of particles associated to different fluctuations produced by the end of inflation. 

Focusing on the case of two-field models where the curvature fluctuation interacts with a massless isocurvature field, we obtain solutions for the Bogoliubov coefficients in different interesting regimes. For instance, we shall be able to obtain exact solutions in the case of small values of the turning rate $\Omega / H \ll 1$, leading us to simple analytical expressions for the number-density of particle species at the end of inflation. For large scales, we find that quanta associated to both the curvature and isocurvature fluctuations have  number-densities given by
\be
    n_{\boldsymbol{p}} =n_{\boldsymbol{p}}^{\rm BD} \bigg( 1 + \frac{ \Omega^2}{2 p^2}   \bigg) .
\ee
This power-law contrasts that of Eq.~(\ref{density-single-field}), which presents logarithmic departures from the Bunch-Davies number-density.  
Our results emphasize the difference between particle production produced by gravitational and non-gravitational couplings and may serve to study phenomena such as reheating and decoherence in the context of multifield inflation.

To proceed, we start in Sec.~\ref{section:Single-field-case} by reviewing particle production in Friedmann-Lema\^itre-Robertson-Walker (FLRW) backgrounds. There we introduce the Bogoliubov coefficients commonly encountered in the study of particle creation in single-field inflation and use them to compute relevant quantities such as the number density of particles. In Sec.~\ref{Sec:QdS}, we focus on quasi-de Sitter spacetime and show that the particle density is modulated by the same spectral index as that of the measured power spectrum of curvature fluctuations. In Sec.~\ref{section:multifield-case}, we adapt the previous methods to the case of two-field models exhibiting non-geodesic trajectories and in Sec.~\ref{sec:Bog-constant-lambda} we derive analytical solutions for Bogoliubov coefficients and the number density of Hamiltonian eigenstates in various regimes. We conclude in Sec.~\ref{Sec:conclusions}.

\subsection{Conventions \& notation}
\vspace{10pt}
We shall work with units such that $\hbar = c = 1$. Fourier transforms are written as $ f_{\K} = \int_{\x} e^{- i {\K} \cdot {\x}} f({\x})$, with $\int_{\x} \equiv \int \de^3 x$, while the inverse is given by $f({\x}) = \int_{\K} e^{ i {\K} \cdot {\x}} f_{\K}$, with $\int_{\K} \equiv \frac{1}{(2 \pi)^3} \int \de^3 k$. We use conformal time $\tau$, defined via $\de t = a \de \tau$, in which case, the FLRW line element reads: $\de s^2 = a^2 \left(- \de \tau^2 + \de x^2 \right)$, with $a(\tau)$ the scale factor. Time derivatives with respect to $\tau$ will be denoted as $()'$.

%%%%%%%%%%%%%%%%%%%%%%%%%%%%%%%%%%%%%%%%%%%%%%%%%%%%%%%%%%%%%%%%%%%%%%%%%%%%%%%%%%%%%%%%%%%%

\setcounter{equation}{0}
\section{Review of FLRW particle production}  
\label{section:Single-field-case}

Working with conformal time, a generic quadratic action describing a fluctuation $\phi (\x,\tau)$, in an FLRW background, has the form
\be
S = \frac{1}{2} \int \de^3 x \de\tau \,f a^2 \left[ ( \phi' )^2 -(\nabla \phi)^2 - a^2 m^2 \phi^2  \right] . \label{action-scalar-1}
\ee
Here, $f(\tau)$ is an arbitrary function of time that may be adjusted as desired, and $m$ is the mass parameter of the field (which may also depend on time). The equation of motion derived from this action reads
\be
\phi'' +  \left(2 a H + \frac{f'}{f}\right)  \phi' -  \nabla^2 \phi + a^2 m^2 \phi = 0 , \label{eq-of-motion} 
\ee
where $H = %\dot a / a =  
a'/a^2$ is the Hubble parameter. From Eq.~(\ref{action-scalar-1}), it follows that the canonical momentum is $\Pi (\x,\tau) = f(\tau) a^2(\tau) \phi' (\x,\tau)$, and that the Hamiltonian of the system is given by
\be
{\cal H} = \frac{f a^2}{2}  \int_\x  \left[ \frac{\Pi^2}{f^2 a^4} +  (\nabla \phi)^2 + a^2 m^2 \phi^2 \right] . \label{hamiltonian-1-field}
\ee 

As usual, upon quantization the fields $\phi$ and $\Pi$ are promoted to operators $\hat \phi$ and $\hat \Pi$ that satisfy the canonical commutation relation (with $\hbar = 1$):
\be
\Big[ \hat\phi (\x , t) ,  \hat\Pi (\y , t) \Big] = i \delta^{(3)} (\x - \y) . \label{can-comm-rel}
\ee
To fulfill this relation, one may introduce creation and annihilation operators $\hat a^\dag_\K$ and $\hat a_\K$ obeying $\left[ \hat a_{\K} , \hat a^\dag_{\q} \right] = (2 \pi)^3 \delta^{(3)} (\K - \q)$, and use them to expand the fields in Fourier space, 
with operator modes given by
\bea
\mqty(  \hat\phi_\K(\tau) \\   \hat\Pi_\K(\tau) )  &=& \mqty( \phi_{k}   & \phi_{k}^*   \\ \Pi_{k}  & \Pi^*_{k}  ) \mqty( \hat a_\K  \\ \hat a^\dag_{-\K} ) . \label{phi-Pi}
\eea
The commutation relation (\ref{can-comm-rel}) holds as long as 
\be
\phi_k (\tau) \Pi^*_{k} (\tau)  - \phi^*_k (\tau) \Pi_{k} (\tau)   = i ,  \label{wronskian-cond-1}
\ee
which happens to be a constant of motion. Let us note that due to this condition, the mode-matrix of Eq.~\eqref{phi-Pi} becomes an element of ${\rm SU}(1,1)$. (Writing Eq.~\eqref{phi-Pi} with $\hat\Pi\to -i\hat\Pi$ renders the determinant equal to one).

In these variables, the equation of motion~(\ref{eq-of-motion}) can be written in the Hamiltonian form,
\begin{align} \label{Hamilton-eqs}
      \mqty(  \phi_k \\   \Pi_k )' = \mqty(0 & \frac{1}{f a^2} \\ - f a^2 \omega_k^2 & 0 ) \mqty( \phi_k \\ \Pi_k ), 
\end{align}
where we have defined $\omega(k,\tau)$ as the standard dispersion relation for a particle in Minkowski spacetime, modulo a scaling introduced by $a(\tau)$ due to the expanding universe:
\be
\omega(k,\tau) = \sqrt{ k^2 + a^2 m^2} . \label{dispersion-relation}
\ee
Promoting the Hamiltonian~(\ref{hamiltonian-1-field}) to an operator and using the expansion~(\ref{phi-Pi}), one obtains
\bea
\hat{\mathcal H} &=& \frac{1}{2} \int_\K \bigg[ A (k,\tau) \hat a_\K  \hat a_{- \K} + A^* (k,\tau) \hat a^\dag_\K  \hat a^\dag_{- \K}   +  B (k,\tau) \left(  \hat a_\K  \hat a^\dag_\K  + \hat a^\dag_\K  \hat a_\K  \right) \bigg] ,  \label{hamiltonian-1-field-quantum}
\eea
where we have defined $A (k,\tau)$ and $B (k,\tau)$ as
\bea
A (k,\tau) &=& \frac{1}{f a^2} \Pi^2_{k} + f a^2 \omega^2(k,\tau) \phi_k^2 ,   \\
B (k,\tau) &=&   \frac{1}{f a^2} | \Pi_{k}  |^2  + f a^2 \omega^2(k,\tau) |\phi_k|^2 .
\eea
Notice that $A$ and $B$ satisfy $B^2 (k,\tau) - | A (k,\tau) |^2 =  \omega^2(k,\tau)$, which may be shown with direct use of Eq.~(\ref{wronskian-cond-1}).

\subsection{Vacuum states and Bogoliubov transformations} \label{section:vacuum-bog} \label{section:intro_of_Bog_trans}

The main tool we will use to compute occupation numbers is the Bogoliubov transformation. We refer to~\cite{Ford:2021syk,Kolb:2023ydq} for recent reviews on their use in the context of cosmological particle creation.
The creation and annihilation operators $\hat a^\dag_\K$ and $\hat a_\K$ introduced above, allow one to count with a definition of a vacuum state $| \Omega \rangle$. This must satisfy
\be
\hat a_\K| \Omega \rangle=0 . \label{vacuum-def}
\ee
With the help of $|\Omega\rangle$ one can construct momentum-eigenstates via operations of the form $\hat a^\dag_{\K_1} \cdots \hat a^\dag_{\K_N} | \Omega \rangle$. Given that the Hamiltonian (\ref{hamiltonian-1-field-quantum}) contains terms proportional to $\hat a^\dag_\K  \hat a^\dag_{- \K}$ and $\hat a_\K  \hat a_{- \K}$ that create and annihilate particle-pairs, $| \Omega \rangle$ is a vacuum state only at a specific time $\tau_0$. This is because the unitary time-evolution operator $\hat U(\tau,\tau_0) \equiv \mathcal{T} e^{- i \int^{\tau}_{\tau_0} \hat{\mathcal{H}} \de \tau}$ (with $\mathcal T$ the time-ordering symbol) does not commute with $\hat a^\dag_\K$ and $\hat a_\K$, and hence $|\Omega(\tau)\rangle = \hat U(\tau,\tau_0) | \Omega \rangle$ does not satisfy (\ref{vacuum-def}) for times $\tau \neq \tau_0$. Thus, as the universe expands, the state $| \Omega \rangle$ becomes populated by particles created due to curvature~\cite{Parker:1968mv}. 

Recall that the definition of $\hat a^\dag_\K$ and $\hat a_\K$ corresponds to a specific form of the mode function $\phi_{k} (\tau)$. Indeed, $\phi_{k} (\tau)$ is the solution of the second order differential equation (\ref{eq-of-motion}) in Fourier space; thus, to completely specify its form one must fix  two integration constants. After imposing the quantization condition (\ref{wronskian-cond-1}) one is left with a single integration constant, whose value defines the vacuum state $| \Omega \rangle$.\footnote{The usual choice within the study of inflation (to be discussed in Sec.~\ref{sec:BD}) leads to the Bunch-Davies state $| \Omega_{\rm BD} \rangle$. For the present discussion, this particular choice is completely irrelevant.} However, after fixing the value of this constant, we may always redefine it without affecting the quantization condition by employing a Bogoliubov transformation. These are transformations that allow us to write new creation and annihilation operators $\hat b^\dag_\K$ and $\hat b_\K$ in terms of $\hat a^\dag_\K$ and $\hat a_\K$ already at hand:
\be
\mqty( \hat b_\K \\ \hat {b}^\dag_{-\K} ) =  \mqty(\alpha^* & -  \beta^* \\ -  \beta & \alpha  ) \mqty( \hat a_{\K} \\ \hat{a}^{\dag}_{-\K} ) ,  \label{bog-trans-1}
\ee
where $\alpha$ and $\beta$ are the Bogoliubov coefficients restricted to satisfy
\be
|\alpha_k(\tau) |^2 - | \beta_k(\tau) |^2 = 1 . \label{bogo-basic-constraint-0}
\ee
This condition ensures that the transformation is canonical such that the new operators $\hat b^\dag_\K$ and $\hat b_\K$ respect the standard commutation relation $\left[ \hat b_{\K} , \hat b^\dag_{\q} \right] = (2 \pi)^3 \delta^{(3)} (\K - \q)$. Note that under~\eqref{bogo-basic-constraint-0}, Bogoliubov transformations span the group ${\rm SU}(1,1)$. See~\cite{Grain:2019vnq} for more details.

If we now expand the field and its conjugate momentum using the new basis, Eq.~(\ref{phi-Pi}) takes the form
\begin{align} \label{b-basis-gen}
 \mqty(\hat \phi_\K(\tau) \\ \hat \Pi_\K(\tau)) &=  \mqty( \bar \phi_k  &  \bar \phi^*_k  \\   \bar \Pi_k  &   \bar \Pi^*_k   ) \mqty(\hat b_\K \\ \hat b^\dag_{-\K}),
\end{align}
where, $\bar\phi_{k} (\tau)$ and $ \bar \Pi_{k} (\tau) $ are new mode functions related ---due to Eq.~\eqref{bog-trans-1}--- to the $\hat a$-basis ones through
\bea
 \mqty(\bar\phi_{k} & \bar\phi^*_{k} \\ \bar\Pi_{k} & \bar\Pi_{k}^* ) &=  \mqty( \phi_k  &  \phi^*_k \\   \Pi_k &  \Pi^*_k ) \mqty(\alpha & \beta^* \\ \beta & \alpha^*).
 \label{phi-bar-Pi}
 \eea
Given that the new mode functions $\bar \phi_{k} (\tau)$ and $\bar \Pi_{k} (\tau)$ satisfy the same quantization condition (\ref{wronskian-cond-1}), it follows that the Bogoliubov transformation is nothing but a redefinition of the second integration constant that determines the form of $\phi_{k} (\tau)$. 

The operators $\hat b^\dag_\K$ and $\hat b_\K$ define a new vacuum state $|\overline{ \Omega}\rangle$ via
\be
\hat b_\K | \overline{ \Omega} \rangle = 0 .
\ee
It should be clear that $| \overline{ \Omega} \rangle$ and $| \Omega \rangle$ have different particle contents, since $\hat b_\K | \Omega \rangle \neq 0$ (that is, $| \Omega \rangle$ contains $b$-particles). In fact, defining the number operator for $b$-particles as $\hat N^{b}_\K \equiv \hat b^\dag_\K \hat b_\K$, it follows that the expectation value of the number density of $b$-excitations in the $|\Omega\rangle$ state, is given by
\be
n_\K \equiv \frac{1}{V} \langle \Omega | \hat N_\K  | \Omega \rangle   = | \beta_k |^2 , \label{number-density}
\ee
where we identified $V = (2 \pi)^3  \delta^{(3)} (\K - \K)$ to cancel the volume $V$ appearing in the denominator of the middle expression.

%%%
%%%
\subsection{Hamiltonian diagonalization in the adiabatic basis}
\label{section:diag-H}

In general, the momentum eigenstates $\hat a^\dag_{\K_1} \cdots \hat a^\dag_{\K_N} | \Omega \rangle$ constructed above are not eigenstates of the Hamiltonian $\mathcal H$. In order to find common eigenstates of both the Hamiltonian and momentum operators, one has to construct a vacuum state $|\Omega_0\rangle$ that allows for such particle states~\cite{Imamura:1960tzx,Parker:1969au} at any given fixed time $\tau_0$. This may be built at $\tau_0$ by performing a Bogoliubov transformation of the form~\eqref{bog-trans-1}:
\be
\mqty( \hat b_\K (\tau_0)\\ \hat {b}^\dag_{-\K} (\tau_0)) =  \mqty(\bar\alpha^*  & -  \bar\beta^*  \\  -  \bar\beta  & \bar\alpha  ) \mqty( \hat a_{\K} \\ \hat{a}^{\dag}_{-\K} ) ,  \label{bogo-diag}
\ee
with $\left|\bar\alpha_k (\tau_0) \right|^2 - | \bar\beta_k (\tau_0) |^2 = 1$, such that in the $\hat b$-basis, 
the Hamiltonian is diagonal:
\begin{equation}  \label{Single-Ham-b-basis}
    \hat {\mathcal H}(\tau_0) = \frac{1}{2} \int_{\K} E_k(\tau_0) \bigg( \hat b^\dag_\K(\tau_0) \hat b_\K(\tau_0) + \hat b_\K(\tau_0) \hat b^\dag_\K( \tau_0) \bigg) ,
\end{equation}
with $E_k(\tau_0)$ its eigenvalue. Let us review one way to obtain the Bogoliubov coefficients $\left(\bar\alpha,\bar\beta \right)$.

\subsubsection{Setting up the eigenvalue problem}
First, we compute $E_k$. In the instantaneous $\hat b$-basis, the field and its conjugate momentum are given by Eq.~\eqref{b-basis-gen} evaluated at $\tau_0$, with the identification $\hat b_\K\equiv\hat b_\K(\tau_0)$. 
The diagonal form of the Hamiltonian implies the commutation relations 
\begin{align} \label{com-diag-H-fi}
   \mqty(\hat \phi_\K(\tau_0) \\ \hat \Pi_\K(\tau_0))' &= i \mqty( \big[\hat{\mathcal H}(\tau_0) , \hat \phi_\K(\tau_0) \big] \\ \big[ \hat{\mathcal H}(\tau_0) , \hat \Pi_\K(\tau_0) \big] ) = iE_k \mqty(  -   \bar \phi_k  &  \bar \phi^*_k \\ -   \bar \Pi_k  &  \bar \Pi^*_k   ) \mqty(\hat b_\K  \\ \hat b^\dag_{-\K} ) ,
\end{align}
where everything in the right-hand side is evaluated at $\tau=\tau_0$.
Upon replacing (\ref{Hamilton-eqs}) into the leftmost part of Eq.~\eqref{com-diag-H-fi}, expanding in the $\hat b$-basis, and matching with the rightmost part,  we arrive at
\begin{align} \label{algebraic-rel-sf}
    \mqty( i E_k & \frac{1}{f a^2} \\ -f a^2 \omega_k^2 & i E_k ) \mqty(\bar \phi_k \\ \bar \Pi_k) = 0 ,
\end{align}
from which we can directly compute the eigenvalue $E_k$. The solution turns out to be the FLRW dispersion relation~\eqref{dispersion-relation}:
\begin{align} \label{dispersion-again}
   E_k^2 =  \omega_k^2 = k^2 + a^2 m^2 .
\end{align}

Having the eigenvalue, we may next compute the $\hat b$-basis mode functions $\left(\bar\phi,\bar\Pi\right)$. These are given as solutions of the algebraic system~\eqref{algebraic-rel-sf} (which degenerates to a single equation) and
the Wronskian condition~\eqref{wronskian-cond-1},
leading to
\begin{align} \label{fi-bar}
   \left|\bar \phi_k \right|^2 = \frac{1}{2 f a^2 \omega_k} \qquad\&\qquad \left|\bar\Pi_k \right|^2= \frac{fa^2\omega_k}{2}.
\end{align}
Note that the field mode-function is determined up to a phase: $\bar \phi_k = \left|\bar \phi_k \right|e^{iS_k}$. Also, from Eq.~\eqref{algebraic-rel-sf}, it follows that $\bar\Pi_k=\left| \bar \Pi_k \right| e^{iS_k-i\pi/2}$.

Next, we compute the Bogoliubov coefficients $\left(\bar\alpha,\bar\beta\right)$ of the transformation~\eqref{bogo-diag}. Inverting Eq.~\eqref{phi-bar-Pi} with $\left(\alpha,\beta\right)\to\left(\bar\alpha,\bar\beta\right)$
and using~\eqref{wronskian-cond-1} and~\eqref{fi-bar}, we find
\begin{align} \label{bogo-diag-1}
    \bar\alpha_k(\tau_0) &= 
    - i \frac{\Pi_k^* + i f a^2 \omega_k \phi_k^*}{\sqrt{2 f a^2 \omega_k} } e^{i S_k(\tau_0)} , 
    \\ \label{bogo-diag-2}
   \bar \beta_k(\tau_0) 
   & = i \frac{\Pi_k + i f a^2 \omega_k \phi_k}{\sqrt{2 f a^2 \omega_k}}  e^{i S_k(\tau_0)}.
\end{align}
With this Bogoliubov transformation we have gained a state $| \Omega_0 \rangle$ that serves as the basis to construct particle states with a dispersion relation (\ref{dispersion-relation}). Just as with $| \Omega \rangle$, the state $| \Omega_0 \rangle$ will become populated as time passes (again, because the unitary operator $\hat U(\tau, \tau_0)$ does not commute with $\hat b_\K$). 

Finally, we may compute the number density $n_k(\tau_0)$ of Hamiltonian eigenstates populating the state $|\Omega\rangle$ originally defined in (\ref{vacuum-def}). Putting together~\eqref{number-density} and~\eqref{fi-bar}, one arrives at~\cite{Mamaev:1976zb}
\begin{align}
    n_k(\tau_0) &= \frac{f a^2}{2 \omega_k}\Big( |\phi_k'|^2 + \omega_k^2 |\phi_k|^2 \Big) - \frac{1}{2}  .
\label{number-density-bog}
\end{align}
Recall that both the mode function $\phi_k$ and the state $|\Omega\rangle$ are determined by a single integration constant (the other one set by the quantization condition). Thus,~\eqref{number-density-bog} informs us that different states lead to different particle contents. In particular, the transformation~\eqref{bogo-diag-1}-\eqref{bogo-diag-2} yields a mode function $\bar \phi_k (\tau)$ for which $ n_\K(\tau_0) = 0$. 

\subsubsection{The adiabatic basis}
Following~\cite{Zeldovich:1971mw}, we may derive dynamical equations for the Bogoliubov coefficients $\left(\bar\alpha,\bar\beta\right)$ that bring us to the diagonal basis at any time $\tau$. We start by parametrizing the diagonal modes with the following ansatz:
\begin{align} \label{diag-modes}
    \bar \phi_k(\tau) = \frac{ e^{i S_k(\tau)}}{\sqrt{2 f a^2 \omega_k}} , \quad \bar \Pi_k(\tau) = - i f a^2 \omega_k \bar \phi_k ,
\end{align}
with $\omega_k$ given by~\eqref{dispersion-again}. 
Next, inverting the Bogoliubov transformation \eqref{phi-bar-Pi}, we obtain
\begin{align}
    \mqty(  \phi_k \\   \Pi_k ) = \mqty(\bar\phi_k & -\bar\phi_k^* \\ \bar\Pi_k & -\bar\Pi_k^* ) \mqty( \bar\alpha^* \\ \bar\beta ) ,
    \label{adiabatic-mode-2}
\end{align}
where we have suppressed the time dependence. Putting together~\eqref{adiabatic-mode-2} with (\ref{Hamilton-eqs}) and using the ansatz~\eqref{diag-modes}, one obtains 
\begin{align}
    \bigg[ \bar\alpha^{*\prime} + \Big(i\omega_k+i S_k'-\frac{1}{2} F_k \Big)\bar\alpha^* \bigg]\bar\phi_k - \bigg[ \bar\beta^{\prime} - \Big(i\omega_k+i S_k'+\frac{1}{2} F_k \Big)\bar\beta \bigg]\bar\phi_k^* &=0 , \\
    \bigg[ \bar\alpha^{*\prime} + \Big(i\omega_k + i S_k'+\frac{1}{2} F_k \Big)\bar\alpha^* \bigg]\bar\Pi_k - \bigg[ \bar\beta^{\prime} - \Big(i\omega_k + i S_k'-\frac{1}{2} F_k \Big)\bar\beta \bigg]\bar\Pi_k^* &=0 ,
\end{align}
where we defined 
\be \label{F-adiab}
F_k = \frac{\left(fa^2\omega_k \right)'}{fa^2 \omega_k}.
\ee 
This system can be recast as a first order system for the Bogoliubov coefficients:
\begin{align} \label{diag-eom}
    \mqty( \bar\alpha^* \\ \bar\beta)' = i \mqty( -\left(\omega_k + S_k' \right) & \frac{i}{2}F_k e^{-2i S_k} \\ \frac{i}{2}F_k e^{2i S_k} & \omega_k + S_k' ) \mqty(\bar\alpha^* \\\bar \beta) .
\end{align}

Up until now, $S_k$ has been an arbitrary function parametrizing the ansatz (\ref{diag-modes}). Different choices of $S_k$ return different results for $\left(\bar\phi_k , \bar \Pi_k\right)$ and $\left(\bar \alpha_k , \bar \beta_k\right)$ without affecting the form of $\left(\phi_k , \Pi_k\right)$. A typical choice consists in $S_k' =-\omega_k$, such that the diagonal modes are given by
\begin{align} \label{adiabatic-mode}
    \bar\phi_k(\tau) = \frac{e^{-i\int^\tau \!\dd\bar\tau \,\omega_k }}{\sqrt{2fa^2 \omega_k}}.
\end{align}
Then, from Eq.~\eqref{adiabatic-mode-2}, the $\hat a$-basis mode function can be written as~\cite{Zeldovich:1971mw,Kofman:1997yn,Nilles:2001fg,Martin:2015qta}
\begin{align}
  \phi_k(\tau) = \bar\alpha_k^*(\tau) \frac{e^{-i\int^\tau  \!\dd\bar\tau \,\omega_k}}{\sqrt{2fa^2 \omega_k}} - \bar\beta_k(\tau) \frac{e^{i\int^\tau  \!\dd\bar\tau \,\omega_k}}{\sqrt{2fa^2 \omega_k}} .
  \label{adiabatic-mode-function}
\end{align}
This is an exact solution to the equation of motion~(\ref{Hamilton-eqs}). By taking the coefficients as constants $\alpha(\tau_0),\beta(\tau_0)$, \emph{i.e.}, fixed by boundary conditions, one obtains the lowest-order adiabatic approximation~\cite{Parker:1969au,Fulling:1979ac,Traschen:1990sw} or WKB solution, which works well whenever $F_k$ in Eq.~\eqref{F-adiab} evolves slowly (adiabatically) or the interval of interest (say between $\tau_1$ and $\tau_2$) is short enough $\int_{\tau_1}^{\tau_2} \dd\tau \, a H  \ll 1 $. 

Note that the adiabatic basis~\eqref{adiabatic-mode-function} differs from the one commonly used in the literature (see \emph{e.g.}~\cite{Zeldovich:1971mw,Martin:2015qta}) in the form of the dispersion relation~\eqref{dispersion-again}. This is due to the fact that many authors prefer to work with canonically normalized fields, which adds a constant mass to the modes set by the Hubble rate. This however restricts the momentum range wherein one can diagonalize the Hamiltonian due to instabilities for modes with $k < H$. In order to avoid such problems, here we work with standard massless scalar modes whose Hamiltonian coincides with the temporal component of the energy momentum tensor~\cite{Casagrande:2023fjk}.

\subsubsection{The Bunch-Davies state}
\label{sec:BD}

As an application of the previous discussion, let us analyze the example of a massless scalar field on a fixed de Sitter background. Here, the scale factor is given by $a= - \frac{1}{H \tau}$ (with $H$ a constant) and the conformal time $\tau$ covers the range $\tau \in (- \infty , 0)$. The action describing a massless canonical scalar field corresponds to (\ref{action-scalar-1}) with the particular values $f=1$ and $m=0$:
\be \label{action-scalar-mless}
S = \frac{1}{2} \int \de^3 x \de \tau a^2 \left[ ({\phi'})^2 -  (\nabla \phi)^2  \right] . 
\ee
The equation of motion in Fourier space reads
\be
\phi_k'' - \frac{2}{\tau} \, \phi_k' + k^2 \phi_k = 0 , \label{eom-de-sitter}
\ee
whose general solution is given by $\phi_k (\tau) = C_1  \left( 1 + i k \tau  \right) e^{- i k\tau} + C_2 \left( 1 - i k \tau  \right) e^{ i k \tau}$, with $C_1$ and $C_2$ integration constants. As already mentioned, one may partially fix their values by imposing the quantization condition~(\ref{wronskian-cond-1}); the remaining freedom then determines the state $| \Omega \rangle$. 

The customary procedure here is to ask that the mode function $\phi_k$ coincide with the mode function of a quantized field in Minkowski spacetime at very short wavelengths ($k / a \gg H$). This step selects positive frequency modes of the form
\be
\phi^{\rm BD}_k (\tau) = \frac{i H}{\sqrt{2k^3}} \left( 1 + i k \tau  \right) e^{- i k \tau } . \label{BD-mode-function}
\ee
The resulting state $|\Omega_{\rm BD} \rangle$ associated to this mode function is the Bunch-Davies state. 

Interestingly, one may interpret the form of (\ref{BD-mode-function}) as resulting from pair creation of massless particles due to the temporal evolution of the flat patch of de Sitter with respect to a Minkowski spacetime. To see this, let us write the BD modes~\eqref{BD-mode-function} in the representation~\eqref{adiabatic-mode-function}. To do so, we may solve the system~\eqref{diag-eom} for $-S_k'=\omega_k=k$, to find\footnote{Equivalently, the expressions~\eqref{al-be} can be obtained by plugging the BD mode function~\eqref{BD-mode-function} into Eqs.~(\ref{bogo-diag-1},~\ref{bogo-diag-2}) and fixing the phase as $S_k(\tau)=-k\tau$.}
\be  \label{al-be}
\bar\alpha_k (\tau) = -\frac{i}{2k\tau} \left(1-2ik\tau\right),\quad \bar\beta_k(\tau) = - \frac{i}{2k\tau} e^{-2ik\tau},
\ee
where we imposed asymptotic conditions $\alpha=1$ and $\beta=0$ in the Minkowski limit $k|\tau|\gg 1$. One may then check that the BD mode function~\eqref{BD-mode-function} can be obtained via mode mixing~\cite{Konieczka:2014zja}, as
\be
\phi^{\rm BD}_k (\tau) =\bar \alpha^*_k(\tau) \frac{e^{-ik\tau}}{\sqrt{2ka^2}} - \bar\beta_k(\tau) \frac{e^{ik\tau}}{\sqrt{2ka^2}}. \label{BD-from-M}
\ee
Note that, by construction, these are the coefficients that diagonalize the Hamiltonian.
Evidently, in the Minkowski limit only the positive mode survives ($1/\sqrt{2ka^2}$ is the canonical normalisation for Minkowski plane waves).

We may now ask what is the particle content of the BD state. Recall that here, by particles we mean energy and momentum eigenstates. The answer is then obtained by replacing (\ref{BD-mode-function}) back into (\ref{number-density-bog}) with $f = 1$. One finds~\cite{Garbrecht:2002pd}
\be 
 n_\K^{\rm BD} (\tau)  = \frac{1}{4(k\tau)^2}. \label{n-BD}
\ee
This result informs us that $|\Omega_{\rm BD} \rangle$ has a non-vanishing particle content that goes to zero at very short wavelengths, in consistency with the requirement that we recover Minkowski space in this limit. It is instructive to also compute the number of particles per physical momentum $p =-H k\tau$. First, we integrate~\eqref{n-BD} over the solid angle and divide by $(2\pi)^3$:
\be 
 n_p^{\rm BD}   = \frac{1}{2} \left(\frac{ H}{2\pi}\right)^2 \frac{1}{p^2}. \label{n-BD-phys}
\ee
Hence, the number-density of states with momentum less or equal to $p$ is $N(p) =  \int_0^{p} \de p\, p^2  n_{p}^{\rm BD} $. It then follows that the number-density of states $\de N_{\rm BD}(p)$ within $p$ and $p+\dd p$ is simply given by
\be  \label{N-phys}
 \de N_{\rm BD}(p)  = \frac{1}{2} \left(\frac{ H}{2\pi}\right)^2 \de p . 
\ee
In other words, the number-density of particles per physical momentum is constant. Recall that, as the universe expands, the physical momentum of a particular state gets redshifted. Therefore, (\ref{N-phys}) reveals that, as time passes, the number of particles exiting a fixed momentum-range is replaced by the same number of particles entering it. This result is consistent with the fact that in de Sitter there is no preferred time. In addition, notice that (\ref{N-phys}) favours no particular momentum~\cite{Gibbons:1977mu}. 
\paragraph{A curious observation} Consider a Bose-Einstein distribution of scalar particles of mass $M$, with chemical potential $\mu$ and temperature $T$:
\be \label{BE}
n(\varepsilon)   = \dfrac12 \frac{1}{e^{ \frac{\varepsilon-\mu}{ T}}-1 },
\ee
with a dispersion relation given by $\varepsilon(p)=\sqrt{p^2 + M^2}$. The $1/2$ overall factor accounts for pair production. Upon setting the temperature to the Gibbons-Hawking value~\cite{Gibbons:1977mu} $T_{\rm dS}\equiv H/(2\pi)$, the mass and the chemical potential equal to $\mu=M=T_{\rm dS}/2$ and taking the non-relativistic limit, $p\to0$,  
one obtains~\eqref{n-BD-phys}. That is, from a superhorizon perspective the distribution~\eqref{n-BD-phys} can be perceived as thermal with $T=T_{\rm dS}$. It would be interesting to understand the nature of this thermal bath and its relation to the Unruh effect~\cite{Davies:1974th,Unruh:1976db}. A similar observation has been reported in~\cite{Markkanen:2016aes}, which only holds for a heavy massive scalar. Here, the scalar field is massless as evident from the action~\eqref{action-scalar-mless}. The mass $M=T_{\rm dS}/2$ appearing in the dispersion relation $\varepsilon(p)$ of Eq.~\eqref{BE} is not that of the scalar field; it is the mass of the apparent bath's particles.

\setcounter{equation}{0}
\section{The spectral index from particle creation in slow-roll inflation} \label{Sec:QdS}

In order to isolate effects on top of the background~\eqref{n-BD} we now proceed to generalize the previous construction to the case where the adiabaticity is broken not by the time dependence induced by the exponentially expanding universe but due to a deviation from an exact de Sitter background. In this case, the ``adiabatic" modes are the BD modes instead of the Minkowski plane waves, the former being now the first approximation (the BD mode function) to the exact solution (the quasi-de Sitter mode function). In particular, in Sec.~\ref{subsec:ns}, we will see how the spectral index encodes the production of these de Sitter particles due to slow-roll inflation.

Slow-roll inflation corresponds to dynamics on a quasi-de Sitter spacetime. The action describing the comoving curvature perturbation $\zeta(\x , t)$ is then given by
\be
S =  \int \de^3 x \de\tau \,\epsilon a^2 \left[ {\zeta'}^2 - (\nabla \zeta)^2  \right] ,
\ee
where $\epsilon = - \dot H / H^2 > 0$ is the first slow-roll parameter. We will also encounter the second slow-roll parameter $\eta \equiv \dot \epsilon / H \epsilon$. Notice that this action corresponds to that of Eq.~(\ref{action-scalar-1}) with the replacements $\phi = \zeta$, $f = 2 \epsilon$ and $m=0$. 
Instead of working directly with $\zeta$, it will be convenient to consider the dynamics of the Mukhanov-Sasaki field:
\be
Q (\x , \tau) \equiv \sqrt{2 \epsilon} a \zeta (\x , \tau) .
\ee
This will allow us to explore the limit $\epsilon \to 0$ more directly.\footnote{As a trade-off, one has to be careful when constructing the adiabatic basis since field redefinitions can lead to different Hamiltonians~\cite{Casagrande:2023fjk}.} Taking into account slow-roll corrections, one may write $a(\tau) = - \frac{1}{(1 - \epsilon) H \tau}$. Beware, here $\epsilon$ and $H$ are functions of time that evolve slowly, and to first order in slow-roll corrections they are given by 
\be \label{eH(t)}
\epsilon (\tau) = \epsilon_0 \left( 1 - \eta_0 \ln H_0 |\tau| \right) \quad \& \quad H (\tau) = H_0 \left( 1 + \epsilon_0 \ln H_0 |\tau| \right),
\ee
where $\epsilon_0$, $H_0$ and $\eta_0$ are the values of $\epsilon$, $H$ and $\eta$ at conformal time $\tau_0 = - 1/H_0$. 

To leading order in slow roll, the action for $Q$ becomes
\be
S = \frac{1}{2} \int \dd[3] x \dd \tau  \left[  (Q')^2  - (\nabla Q)^2 + \frac{1}{ \tau^2}   \Big[ 2 - \frac{3}{2} (n_s - 1)  \Big]  Q^2 \right] , \label{Action-Q}
\ee
where $n_s - 1 = - 2 \epsilon_0 - \eta_0$. Expanding the Fourier duals of  $Q(\x , \tau)$ and its momentum $P(\x , \tau)$ in the $\hat a$-basis as in Eqs.~(\ref{phi-Pi}), we are led to the following equation of motion:
\be
P'_{k} + k^2 Q_k - \frac{1}{ \tau^2}   \Big[ 2 - \frac{3}{2} (n_s - 1)  \Big]  Q_k = 0, \qquad P_{k} = Q_k' . \label{Q-eq} 
\ee
As usual, the mode functions $Q_k(\tau)$ and $P_{k}(\tau)$ must satisfy the quantization condition (\ref{wronskian-cond-1}) with the replacements  $\phi_k \to Q_k$ and $\Pi_k \to P_{k}$.

\subsection{Quasi-de Sitter dynamics as a Bogoliubov transformation} \label{time-dependent-Bog-single-field}

In the particular case $\epsilon = \eta=0$ we recover de Sitter spacetime with $H$ a constant. Consequently, $Q$ satisfies the equation of motion (\ref{eom-de-sitter}) and we may choose the mode function as the standard Bunch-Davies solution:
\be \label{Q-BD}
Q_{\rm BD}(\tau) =  \frac{1}{\sqrt{2 k}} \left( 1 - \frac{i}{k \tau} \right) e^{- i k \tau } .
\ee
Hereafter, we suppress the momentum labels in the mode functions.
We can use $Q_{\rm BD}$ to obtain the full solution $Q$ to Eq.~(\ref{Q-eq}) through a canonical transformation. To achieve this, let us introduce a new set of time-dependent coefficients $\alpha_k(\tau)$ and $\beta_k(\tau)$ satisfying $|\alpha_k (\tau)|^2 - |\beta_k (\tau)|^2 = 1$ and define the following Bogoliubov transformation:
\bea
\mqty( \hat c_\K  (\tau) \\ \hat c^\dag_{-\K}  (\tau))  &=& \mqty( \alpha   & \beta^*  \\  \beta   & \alpha^* ) \mqty(\hat a_\K \\ \hat a^\dag_{-\K} ) .  \label{a-b-3}
\eea
Here, $\hat c_\K^\dag (\tau)$ and $\hat c_\K( \tau)$ are new creation and annihilation operators satisfying the standard algebra. 

Next, we may expand the mode functions in this basis:
\be
\mqty( Q \\ P  ) = \mqty( Q_{{\rm BD}}  & Q^*_{{\rm BD}}  \\ P_{\rm BD}  & P_{\rm BD}^*  )  \mqty(\alpha  \\ \beta ) . \label{Q-alpha-beta} 
\ee
Note that this equation is equivalent to (\ref{adiabatic-mode-2}) relating the full solution $\phi$ with the adiabatic solution $\bar \phi$, with $Q_{\rm BD}$ playing the same role as $\bar \phi$.
Eq.~(\ref{Q-alpha-beta}) automatically ensures that the pair $(Q,P)$ satisfy canonical commutation relations. However, given that $P = Q'$ and $P_{\rm BD} = Q'_{{\rm BD}}$, it follows that
\be
Q_{\rm BD} (\tau) \alpha_k' (\tau)   + Q_{\rm BD}^* (\tau) \beta_k' (\tau)   = 0 . \label{cond_ab_1}
\ee
Let us also not forget that $Q(\tau)$ must respect the equation of motion (\ref{Q-eq}). This implies a further condition on $\alpha_k (\tau)$ and $\beta_k (\tau)$ and their time derivatives:
\be
P_{\rm BD} (\tau)  \alpha_k' (\tau)     +  P_{\rm BD}^* (\tau)   \beta_k' (\tau) + \frac{3}{2 \tau^2} (n_s - 1) \bigg[ \alpha_k (\tau)  Q_{\rm BD} (\tau) + \beta_k (\tau)  Q_{\rm BD }^* (\tau)\bigg] = 0 . \label{cond_ab_2}
\ee

Putting together Eqs.~(\ref{cond_ab_1}) and~(\ref{cond_ab_2}) while imposing the condition $Q  P^*  -  Q^*  P   = i $, we finally find the following first-order set of differential equations 
for $\alpha_k (\tau)$ and $\beta_k (\tau)$:
\be
\frac{1}{k} \left(\begin{array}{c}\alpha_k \\ \beta_k\end{array}\right)' = \mathcal M (k \tau) \left(\begin{array}{c}\alpha_k \\ \beta_k\end{array}\right) ,  \label{eq-alpha-beta-1}
\ee
with the matrix $\mathcal M ( k \tau)$ given by
\be
\mathcal M (k \tau) = - \frac{3 i}{2 k\tau^2} (n_s - 1) \left[ \begin{array}{cc} | Q_{\rm BD} |^2 &  Q_{\rm BD }^{*2}   \\  -  Q_{\rm BD }^{2 }   &  - | Q_{\rm BD} |^2 \end{array}\right] .
\ee
Equation~(\ref{eq-alpha-beta-1}) is just the analog of (\ref{diag-eom}) in the new bases where $Q_{\rm BD}$ takes the role of the adiabatic mode. Writing  $\mathcal M ( k \tau)$ explicitly in terms of $k \tau$, one finds
\be \label{M(kt)}
\mathcal M (k \tau)= - \frac{3 i}{4 k^2 \tau^2} (n_s - 1) \left[ \begin{array}{cc} 1 + \frac{1}{k^2 \tau^2}   &  \left( 1 + \frac{i}{k \tau} \right)^2 e^{ 2 i k \tau }  \\  - \left( 1 - \frac{i}{k \tau} \right)^2 e^{- 2 i k \tau }   &  - 1 - \frac{1}{k^2 \tau^2} \end{array}\right] .
\ee
It may be verified that Eq.~(\ref{eq-alpha-beta-1}) preserves the condition $|\alpha_k (\tau)|^2 - |\beta_k (\tau)|^2 = 1$. 

The formal solution to this equation with initial conditions $\left(\alpha_k\;\beta_k\right)_{\rm ini} $, can be written with the help of the Dyson series as
\be
\left(\begin{array}{c}\alpha_k \\ \beta_k\end{array}\right) = \mathcal T \exp \left\{ \int^{k \tau}_{-\infty } \!\!\!  \de x \, \mathcal M  (x) \right\} \left(\begin{array}{c}\alpha_k \\ \beta_k \end{array}\right)_{\rm ini} , \label{sol-Dyson-1}
\ee
where $\mathcal T$ is the time ordering symbol.  
The Bunch-Davies initial condition set at $k\tau \to - \infty$, is $\left(\alpha_k\;\beta_k\right)_{\rm ini} = (1\;0)$. We will call the solution~(\ref{sol-Dyson-1}) with BD initial conditions, the quasi-de Sitter mode function:
\be
Q_{\rm qdS} (\tau) = \alpha_k (\tau) Q_{\rm BD}(\tau) + \beta_k (\tau) Q^*_{\rm BD}(\tau) , \qquad  \lim_{k\tau \to - \infty} \left(\begin{array}{c} \alpha_{k}(\tau) \\ \beta_{k} (\tau) \end{array}\right)  \to \left(\begin{array}{c}1 \\0\end{array}\right) . \label{QdS-mode-function}
\ee
This mode function defines the vacuum state $| \Omega_{\rm qdS} \rangle$ satisfying $\hat a_{\K} | \Omega_{\rm qdS} \rangle = 0$.

\subsection{Quasi-de Sitter mode function at first order}
\label{subsec:ns}
Instead of pursuing the integration of the full solution (\ref{sol-Dyson-1}) let us expand the series up to first order in $n_s - 1$ (after all, the action~(\ref{Action-Q}) is valid up to first order in slow-roll corrections). Keeping leading terms in the long wavelength limit ($k |\tau| \ll 1$) one finds:\footnote{As a check, one may verify that these expressions can be directly obtained from Eq.~\eqref{Q-eq}: First one can solve \eqref{Q-eq} for arbitrary constant $n_{\rm s}$, fix the integration constants by imposing that for $n_{\rm s}=1$ the mode function take the form~\eqref{Q-BD}, and then expand the solution to first order in $n_{\rm s}-1$ and identify the coefficients of the BD mode function and its complex conjugate. The outcome (in the soft limit) is indeed~\eqref{sol-alpha-single} and~\eqref{sol-beta-single}.}
\bea
\alpha_k (\tau) &\simeq& 1+ \frac{3 i}{4 } (n_s - 1) \bigg[ \frac{1}{k \tau} + \frac{1}{3 k^3 \tau^3} \bigg] ,
\label{sol-alpha-single}
\\
\beta_k (\tau) &\simeq&  \frac{3 i}{4 } (n_s - 1) \bigg[   \frac{1}{k \tau} + \frac{1}{3 k^3 \tau^3} + \frac{2}{3} i  \ln |k \tau|  \bigg] \label{sol-beta-single} .
\eea
Notice that the solutions~(\ref{sol-alpha-single}) and~(\ref{sol-beta-single}) contain terms that quickly become much larger than $1$ even for small values of $n_s - 1$. However, this is not problematic for the perturbative scheme consisting in keeping the leading term of the Dyson series: for $k|\tau| \ll 1$ the matrix~\eqref{M(kt)} becomes 
\be 
\lim_{k\tau\to0} {\cal M}(k\tau) =  \frac{3 i}{4} \left(\frac{ 1}{k^2 \tau^2} + \frac{1}{k^4 \tau^4}\right) (n_s - 1) \left[ \begin{array}{cc}  1 &  -1  \\  1  &  - 1 \end{array}\right] ,
\ee
and the asymptotic system can be solved exactly to all orders in $n_s-1$. One finds:
\be
\lim_{k\tau\to0} \alpha_k (\tau)-1 = \lim_{k\tau\to0} \beta_k (\tau) =  \frac{3 i}{4 } (n_s - 1) \bigg[ \frac{1}{k \tau} + \frac{1}{3 k^3 \tau^3} \bigg] .
\ee
This implies that the pieces proportional to $ \frac{1}{k \tau} + \frac{1}{3 k^3 \tau^3} $ in Eqs.~\eqref{sol-alpha-single} and~\eqref{sol-beta-single}, are part of the exact asymptotic solution on long wavelengths, whereas the logarithmic piece corresponds to the correction that must remain small.

Having (\ref{sol-alpha-single}) and (\ref{sol-beta-single}), and using~(\ref{Q-alpha-beta}), we can now obtain an expression for $\zeta_k(\tau)$ (and similarly for $\Pi_{\zeta}$) in the long-wavelength limit. This is found to be
\be
\zeta_k (\tau) = - i   \frac{H}{\sqrt{4 \epsilon k^3}} \left[ 1 +  \frac{1}{2 } (n_s - 1)    \ln |k \tau|  \right] .  \label{zeta-tau}
\ee
Then, if we assume that the universe is in the state $|\Psi\rangle = | \Omega_{\rm qdS} \rangle$, by replacing (\ref{zeta-tau}) back in Eq.~(\ref{number-density-bog}) we are finally led to 
\be
 n^{\rm qdS}_\K (\tau)  =    n^{\rm BD}_\K (\tau) \, \big| k \tau \big|^{n_s - 1}   ,  \label{n-QDS}
\ee
with $n^{\rm BD}_\K (\tau)$ corresponding to the Bunch-Davies number density given in~\eqref{n-BD}.

This result may be compared to the power spectrum $P_\zeta(k)$, defined as
\be
\langle \Psi |\hat\zeta_\K (\tau) \hat\zeta_\q( \tau) | \Psi \rangle \equiv (2 \pi)^3 \delta (\K + \q) P_\zeta(k , \tau). \label{power-spectrum-def}
\ee
Assuming a state $|\Psi\rangle = | \Omega_{\rm qdS} \rangle$ and the basis $\left(\hat a,\hat a^\dag\right)$, one finds $P_\zeta(k , \tau) = |\zeta_k (\tau)|^2$. Notice that the evolution of $\epsilon$ and $H$ given by Eq.~\eqref{eH(t)} cancels the logarithm in Eq.~(\ref{zeta-tau}), leading to 
\be
\zeta_k =  - i   \frac{H_0}{\sqrt{4 \epsilon_0 k^3}} \left[ 1 +  \frac{1}{2 } (n_s - 1)    \ln \frac{k}{H_0}  \right], \label{zeta-freeze}
\ee
which is consistent with the fact that $\zeta_k$ must freeze on superhorizon scales. From here follows the well-known result
\be
P_\zeta(k)  = \frac{H_0^2}{4 \epsilon_0 k^3} \left( \frac{k}{H_0}   \right)^{n_s - 1} . \label{Power-zeta}
\ee

In summary, we see that the spectral index determining the scale dependence of the power spectrum (\ref{Power-zeta}), also modulates the momentum dependence of the particle number-density $n_\K (\tau)$ and can be understood as a consequence of de Sitter particle production due to slow-roll dynamics:
\be
n_s - 1 =  \frac{\de }{\de \ln  k|\tau|} \ln \frac{n^{\rm qdS}_\K(\tau) }{ n^{\rm BD}_\K(\tau)} .   
\label{n_BD-QDS}
\ee
Thus, given that a Bunch-Davies state displays a number of particles per physical momentum remaining constant over time, we see that genuine particle production requires departures from exact de Sitter conditions, such as those present in slow-roll inflation (quasi de Sitter) scenarios. Consequentially, particle production in quasi de Sitter is found to be related to the spectral index, which is a physical observable~\cite{Planck:2018jri,Planck:2018vyg}. 

In the previous discussion, we started with a quasi-de Sitter spacetime parametrized by $\epsilon$ and $\eta$, and we recovered the well-known result~\eqref{Power-zeta} that the scale dependence of the power spectrum is determined by the spectral index $n_s-1 = - 2 \epsilon - \eta$. This was obtained thanks to the standard assumption that the state of the universe $| \Psi \rangle$ is set by $| \Omega_{\rm qdS} \rangle$ which, in turn, is determined by the mode function $Q_{\rm qdS} (\tau)$ defined in (\ref{QdS-mode-function}). Suppose, instead, that we had assumed that the state of the universe $| \Psi \rangle$ coincided with $| \Omega_{\rm BD} \rangle$ defined via $\hat{c}_{\K} (\tau_{0}) | \Omega_{\rm BD} \rangle = 0$, where $\tau_{0}$ is a specific time at which we are interested in evaluating observables (for instance, the end of inflation). Then, the definition (\ref{power-spectrum-def}) would have led us to $P_\zeta(k,\tau_0) = |\zeta_{\rm BD} (\tau_0)|^2$, with $\zeta_{\rm BD} (\tau_0) = Q_{\rm BD} (\tau_0) / \sqrt{2 \epsilon} a$. This would have implied a scale invariant power spectrum:
\be
 P_\zeta(k , \tau_0) = \frac{H_0^2}{4 \epsilon_0 k^3} . \label{dS-from-QdS}
\ee
In other words, even if the background spacetime has a slow-roll evolution parametrized by $\epsilon$ and $\eta$, the scale dependence implied by slow-roll corrections would cancel out by the specific particle content of $| \Omega_{\rm BD} \rangle $. 

Conversely, we could consider a de Sitter spacetime for which $\epsilon = \eta = 0$,\footnote{Given that we are generally interested in the power spectrum of $\zeta$, the reader might prefer to consider a quasi-de Sitter spacetime with values of $\epsilon$ and $\eta$ such that $n_s = 1$. This reconsideration will not change the main point behind the statement that follows.} and choose the Bunch-Davies mode function $Q_{\rm BD}(\tau)$ as the solution to the equation of motion~(\ref{eom-de-sitter}) for a massless scalar field. Then, we are free to consider the time-dependent Bogoliubov-transformation (\ref{a-b-3}) but, this time, writing $Q_{\rm BD}(\tau)$ in terms of $Q_{\rm qdS} (\tau)$ as
\bea
Q_{\rm BD} (\tau) &=& \alpha_k (\tau) Q_{\rm qdS} (\tau) + \beta_k (\tau) Q_{\rm qdS}^*(\tau) , \label{Q-alpha-beta-2} \\
Q_{\rm BD}' (\tau) &=& \alpha_k (\tau) Q'_{\rm qdS} (\tau) + \beta_k (\tau) Q_{\rm qdS}^{*\prime}(\tau) , \label{Q'-alpha-beta-2}
\eea
and impose initial conditions $\alpha_k (\tau) =1$ and $\beta_k (\tau) =0$ for subhorizon scales $k |\tau| \gg 1$. In this way, by imposing that the state of the universe corresponds to $|\Omega_{\rm qdS} \rangle$ as determined by the mode function $Q_{\rm qdS} (\tau)$, we end up with a scale dependent spectrum with a spectral index given by $n_s -1 = - 2 \epsilon_0 - \eta_0$. 

To conclude, canonical transformations allow one to trade geometrical features of spacetime, such as the Hubble flow, for algebraic properties of the Hilbert space, like the evolution of Bogoliubov coefficients parametrizing particle production, and vice versa.

Finally, let us note that what we have just encountered is a generic result connecting particle production to scale dependent features of the power spectrum. For example, consider two different states $|\Omega_1\rangle$ and $|\Omega_2\rangle$ with corresponding mode functions related by a time dependent Bogoliubov transformation
\be 
\mqty(\phi_2  \\ \Pi_2 ) = \mqty(\phi_1  & \phi_1^* \\ \Pi_1  & \Pi_1^*  ) \mqty(\alpha  \\ \beta  ) .
\ee 
We may directly use this expansion in the definition of the power spectrum and the number density of eigenstates. The left-hand side gives the observables for $|\Omega_2\rangle$, while the precise combination of Bogoliubov coefficients on the right-hand side allows us to compute observables for $|\Omega_1\rangle$ with the corresponding correction 
\begin{align}
    P_{\phi_2}(k,\tau) &= P_{\phi_1}(k,\tau)\bigg( 1 + 2\Delta P(k,\tau) \bigg) , \\
    n_{\phi_2}(k,\tau) &= n_{\phi_1}(k,\tau) \bigg( 1 + 2\Delta n(k,\tau) \bigg) + \Delta n(k,\tau),
\end{align}
where
\begin{align}
    \Delta P(\tau,k) &= \Re\bigg[\Big(\frac{\phi_1^2}{|\phi_1|^2}\alpha_k + \beta_k \Big)\beta_k^*\bigg] , \\
    \Delta n(\tau,k) &= \Re\bigg[\left(\frac{(\phi_1')^2+\omega_k^2 \phi_1^2}{|\phi_1'|^2+\omega_k |\phi_1|^2}\alpha_k + \beta_k \right)\beta_k^*\bigg] .
\end{align}
For an expanding universe, the Wronskian condition $\phi_1 \phi_1^{\prime *} - \phi_1^* \phi_1^\prime = i/(f_1 a^2)$ implies that at late times we can write
\begin{align}
    |\phi'_1|^2 + \omega_k^2 |\phi_1|^2 \rightarrow \left( (\phi_1')^2 + \omega^2 \phi_1^2 \right) \frac{\phi_1^*}{\phi_1}.
\end{align}
Hence, the corrections to the power spectrum and particle number coincide on superhorizon scales, $\Delta P \sim \Delta n \sim \Delta$, such that 
\begin{align} \label{P2P1}
    \lim_{k\tau\rightarrow 0 } \frac{P_{\phi_2}(k,\tau)}{P_{\phi_1}(k,\tau)} = 
    \lim_{k\tau\rightarrow 0 }\frac{n_{\phi_2}(k,\tau)}{n_{\phi_1}(k,\tau)}  = 1+2\Delta.
\end{align}
In the case  $\phi_2 = \zeta$, $\phi_1 = \zeta_{\rm BD}$, $f = 2 \epsilon$, $|\Omega_0\rangle=|\Omega_{\rm BD}\rangle$ and $|\Omega\rangle=|\Omega_{\rm qdS}\rangle$, we obtain the relation~\eqref{n_BD-QDS}.

\paragraph{Another curious observation}
Interestingly, just like~\eqref{n-BD}, the particle density~\eqref{n-QDS} can also be matched to a thermal distribution with $T=T_{\rm dS}$ in the superhorizon limit. That is, for $p\to0$, and to leading order in $n_s-1$, the density~\eqref{n-QDS} matches a Bose-Einstein distribution with the same values as~\eqref{BE}, \emph{i.e.} $\mu=M=T_{\rm dS}/2$, albeit with an anomalous dispersion relation: $\varepsilon(p)=\sqrt{p^{3-n_s} + M^2}$.

%%%%%%%%%%%%%%%%%%%%%%%%
%%%%%%%%%%%%%%%%%%%%%%%%
\setcounter{equation}{0}
\section{Particle production in multifield inflation} \label{section:multifield-case}

We are now in a good condition to analyze how multiparticle states are created. We will specialize this discussion to the case of two-field models, which will be sufficiently generic to understand the excitation of states in more general setups. The multifield canonical evolution in the adiabatic basis has been considered in~\cite{Nilles:2001fg}, while Ref.~\cite{Konieczka:2014zja} was ---to our knowledge--- the first paper to study particle production from a turning trajectory. Here\footnote{By a rotation in the field space the kinetically coupled EFT~\eqref{multi-field-action} can be brought in the potentially coupled action of~\cite{Nilles:2001fg}.}, following the single-field discussion, we will consider the canonical evolution in the BD basis, in order to isolate effects stemming solely from the field-space structure, that is, moding out the Hubble flow. This allows for a controlled order-by-order solution where the slow-roll corrections decouple from the multifield dynamics.

To start with, in two-field inflation the quadratic action describing the dynamics of linear perturbations takes the form~\cite{GrootNibbelink:2001qt}
\be
S  = \int \! \de^3 x \de\tau \, a^2 \left[ \epsilon \left( \zeta'  - \sqrt{2/\epsilon} \, \Omega a \psi \right)^2 - \epsilon (\nabla \zeta)^2 + \frac{1}{2}  {\psi'}^2 - \frac{1}{2} (\nabla \psi)^2 - \frac{1}{2} a^2  \mu^2 \psi^2  \right] , \label{multi-field-action}
\ee
where $\zeta$ is the primordial comoving curvature perturbation and $\psi$ is the isocurvature perturbation parametrizing fluctuations orthogonal to the inflationary trajectory in field space. In the previous expression, $\mu$ is the entropy mass\footnote{Upon expanding the square in the kinetic term of the action~\eqref{multi-field-action}, one finds the expected $\Omega\zeta'\psi$ coupling characteristic of a turning trajectory. It is in this form however that one can correctly identify the entropy mass $\mu$, i.e., the pole of the $\psi$-propagator~\cite{Achucarro:2016fby}.} and $\Omega$ the turning rate of the background inflationary trajectory. 
The equations of motion derived from~(\ref{multi-field-action}) read
\bea
(D_\tau  \zeta)'  + a H (2 + \eta)  D_\tau   \zeta - \nabla^2   \zeta =0  , \label{eom-1} \\
\psi'' + 2 a H \psi' -  \nabla^2  \psi + a^2 \mu^2   \psi+ a H \sqrt{2 \epsilon}  \lambda D_\tau   \zeta = 0 , \label{eom-2} 
\eea
where, for convenience, we have defined the dimensionless coupling $\lambda$ as
\be
 \lambda(\tau) \equiv \frac{ 2\Omega(\tau)}{H} .
\ee 
In addition, the covariant derivative $D_\tau \zeta$ appearing in Eqs.~(\ref{eom-1},~\ref{eom-2}) corresponds to the particular combination
\be
D_\tau \zeta \equiv   \zeta'  - a H \frac{\lambda}{\sqrt{2 \epsilon}} \psi . \label{cov-der-zeta-def}
\ee
The set of equations (\ref{eom-1}) and (\ref{eom-2}) informs us how $\zeta$ and $\psi$ interact at linear order through the coupling $\lambda$, which may change in time arbitrarily depending on how the background trajectory evolves in field space. 

The canonical momenta $\Pi_\zeta$ and $\Pi_\psi$ associated to the fields $\zeta$ and $\psi$ implied by the form of the action~(\ref{multi-field-action}) are given by
\be
\Pi_\zeta = 2 a^2 \epsilon D_\tau \zeta \,,\quad   
\Pi_\psi = a^2 \psi ' , \label{canonical-mom-zeta-psi-0}
\ee
such that the field and momentum operators satisfy the standard equal-time commutation relations
\bea
\Big[ \hat\zeta(\tau, {\x}) , \hat\Pi_\zeta(\tau,{\y}) \Big] = i \delta^{(3)} ({\x} - {\y}) , \label{commut-fields-1} \\  \Big[ \hat\psi(\tau, {\x}) , 
\hat\Pi_\psi(\tau, {\y}) \Big] = i \delta^{(3)} ({\x} - {\y}) , \label{commut-fields-2}
\eea
with any other commutator vanishing. Finally, the Hamiltonian of the system reads
\be
\mathcal H \!=\!  \int_\x   \left[  \frac{1}{4 a^2 \epsilon} \Pi_\zeta^2 +  a^2 \epsilon (\nabla \zeta)^2 + \frac{1}{2a^2}  \Pi_\psi^2 + \frac{a^2}{2} (\nabla \psi)^2 + \frac{1}{2} a^4  \mu^2 \psi^2 +    \frac{a H}{2} \frac{\lambda}{\sqrt{2 \epsilon}} \left( \Pi_\zeta  \psi + \psi  \Pi_\zeta \right) \right] . \label{Hamiltonian-multi}
\ee

Given that we are dealing with a coupled system of equations, we must expand $\hat\zeta(\tau, {\x})$ and $\hat\psi(\tau, {\x})$ (and their respective momenta) in ladder operators in Fourier space, as
\be \label{multi-Fourier}
\mqty (\hat\zeta_{\K} (\tau) \\ \hat\psi_{\K} (\tau) \\ \hat\Pi_{\zeta_{\K}} (\tau)  \\ \hat\Pi_{\psi_{\K}} (\tau) )  = \mqty (\zeta_1  & \zeta_2  & \zeta_1^*  & \zeta_2^*  \\ \psi_1  & \psi_2  & \psi_1^*  & \psi_2^*   \\ \Pi_1^{\zeta}   & \Pi_2^{\zeta}   & \Pi_1^{\zeta*}  & \Pi_2^{\zeta*}     \\ \Pi_1^{\psi}   & \Pi_2^{\psi}  & \Pi_1^{\psi*}   & \Pi_2^{\psi*}   ) \mqty (\hat a_1 ({\K}) \\ \hat a_2 ({\K}) \\ \hat a^\dag_1 (-{\K})  \\ \hat a^\dag_2 (-{\K}) ), 
\ee 
where the pairs $\hat a^\dag_{a} ({\K})$ and $\hat a_{a} ({\K})$ (with $a=1,2$) are creation and annihilation operators satisfying
\be
\left[ \hat a_{a} ({\K}) , \hat a_{b}^\dag ({\K}') \right] = (2 \pi)^3 \delta_{ab} \delta^{(3)} ({\K} - {\K}') . \label{commutation-a_b}
\ee
In the previous expressions, $\zeta_a $, $\psi_a $ are the mode functions, whereas  $ \Pi_b^{\zeta}    = 2 a^2 \epsilon  D_\tau \zeta_b  $ and $ \Pi_b^{\psi}   = a^2 \psi_b' $ are their associated momenta. These mode functions satisfy the following equations of motion:
\bea
 \frac{\de}{\de\tau}  \Pi^{\zeta}_a   + 2 \epsilon a^2 k^2   \zeta_a &=&0  ,  \qquad     \Pi_a^{\zeta} =   2 a^2 \epsilon  D_\tau \zeta_a , \label{eom_b-1} \\
 \frac{\de}{\de\tau}   \Pi^{\psi}_a  + a^2 \left(  k^2  + a^2 \mu^2 \right)  \psi_a + a H \frac{\lambda}{\sqrt{2 \epsilon}}  \Pi^{\zeta}_a &=& 0 ,     \qquad   \Pi_a^{\psi} = a^2 \psi_a' .\label{eom_b-2} 
\eea 
Inserting~(\ref{multi-Fourier}) back in the Hamiltonian (\ref{Hamiltonian-multi}), one finds
\be
\hat{\mathcal H} = \frac{1}{2} \sum_{b c}   \int_\K  \bigg[ {\cal A}_{bc}  \hat a_b ({\K})  \hat a_c (-{\K})  +  {\cal B}_{bc}  \hat   a^\dag_b ({\K})   \hat a_c ({\K})  +  {\cal B}_{bc}^*  \hat a_b ({\K})  \hat a^\dag_c ({\K})  +  {\cal A}_{bc}^*   \hat a^\dag_b ({\K})  \hat a^\dag_c (-{\K})  \bigg]  \label{Ham-multi-AB} ,
\ee
with ${\cal A}_{ab}$ and ${\cal B}_{ab}$ given by
\bea
{\cal A}_{ab} &=&    \frac{ \Pi^\zeta_a  \Pi^\zeta_b }{2 a^2 \epsilon}  +     \frac{ \Pi^\psi_a  \Pi^\psi_b}{ a^2 }  + 2 a^2 \epsilon k^2  \zeta_a  \zeta_b  +  a^2  \left( k^2 + a^2 \mu^2 \right)  \psi_a  \psi_b + \frac{a H \lambda}{ \sqrt{2 \epsilon}} \left(  \Pi^\zeta_a   \psi_b +   \psi_a \Pi^\zeta_b \right)   ,   \\
{\cal B}_{ab} &=&    \frac{ \Pi^{\zeta *}_a  \Pi^\zeta_b}{2 a^2 \epsilon}   +     \frac{ \Pi^{\psi *}_a   \Pi^\psi_b}{a^2 }  + 2 a^2 \epsilon k^2  \zeta^*_a  \zeta_b  + a^2  \left( k^2 + a^2 \mu^2 \right) \psi_a^*  \psi_b +  \frac{a H \lambda}{\sqrt{2 \epsilon}}  \left(  \Pi^{\zeta *}_a   \psi_b +   \psi^*_a \Pi^{\zeta}_b \right)  . \qquad
\eea
These matrices satisfy ${\cal A}^t = {\cal A}$ and ${\cal B}^\dag = {\cal B}$.

Finally, as in the single-field case, in the canonical formulation it is straightforward to unveil the symplectic structure of the Hilbert space. First, we may observe that the mode-function matrix ${\cal T}_m$ of Eq.~\eqref{multi-Fourier}, preserves the symplectic form:
\be \label{symp}
{\cal T}_m^t \mathbb{\Omega} {\cal T}_m =  \mathbb{\Omega}, \quad{\rm where} \quad \mathbb{\Omega}=\mqty(0 & \mathbbm{1}_{2} \\ -\mathbbm{1}_{2} & 0),
\ee
with $\mathbbm{1}_2$ the two-by-two identity matrix. Thus ${\cal T}_m \in {\rm Sp}(4,\mathbb{C}).$ Next, we need to consider the constraints imposed by the commutation relations~(\ref{commut-fields-1}) and~(\ref{commut-fields-2}).
Writing the mode-matrix in block form, ${\cal T}_m =\mqty ( {\cal Z}  &  {\cal Z}^* \\ {\cal P}  & {\cal P}^*   )$, these constraints read
\bea 
{\cal Z}{\cal P}^\dag - {\cal Z}^*{\cal P}^t &=& i \mathbbm{1}_2,
\\
{\cal Z}{\cal Z}^\dag - {\cal Z}^*{\cal Z}^t = {\cal P}{\cal P}^\dag - {\cal P}^*{\cal P}^t &=& 0.
\eea
These are the defining relations of the symplectic group over the reals; we thus conclude that ${\cal T}_m \in {\rm Sp}(4,\mathbb{R}).$ Note that this reduces correctly to the single-field case: ${\rm Sp}(2,\mathbb{R}) \simeq {\rm SU}(1,1)$~\cite{Grain:2019vnq}, while, on the other hand, it generalizes to ${\cal T}_m \in {\rm Sp}(2n,\mathbb{R})$ for $n$ interacting scalars.

\subsection{The adiabatic basis}
\label{Sec:mf-adiab}

Let us now diagonalize the Hamiltonian to compute the multifield energy/momentum eigenstates. Following the spirit of Ref.~\cite{Nilles:2001fg}, we may repeat the steps of the single-field case, and consider a Bogoliubov transformation of the form
\be
\mqty( \hat b_{\bar 1}(\K, \tau_0) \\  \hat b_{\bar 2}(\K, \tau_0) \\ \hat b^\dagger_{\bar 1}(-\K, \tau_0) \\ \hat b^\dagger_{\bar 2}(-\K, \tau_0)) = \mqty( \bar\alpha^*_{\bar 1 1}  & \bar\alpha^*_{\bar 1 2}  & -\bar\beta^*_{\bar 1 1}  & -\bar\beta^*_{\bar 1 2}  \\  \bar\alpha^*_{\bar 2 1}  & \bar\alpha^*_{\bar 2 2}  & -\bar\beta^*_{\bar 2 1}  & -\bar\beta^*_{\bar 2 2}  \\ -\bar\beta_{\bar 1 1}  & -\bar\beta_{\bar 1 2}  & \bar\alpha_{\bar 1 1}  & \bar\alpha_{\bar 1 2}  \\ -\bar\beta_{\bar 2 1}  & -\bar\beta_{\bar 2 2}  &  \bar\alpha_{\bar 2 1}  & \bar\alpha_{\bar 2 2}  )  \mqty( \hat a_{ 1}(\K) \\  \hat a_{2}(\K) \\ \hat a^\dagger_{1}(-\K) \\ \hat a^\dagger_{2}(-\K)),
\label{multifield-diag-bogo-c}
\ee
constrained to satisfy
\bea
\sum_a \Big(  \alpha_{\bar a a}^* \alpha_{\bar b a} -  \beta_{\bar a a}^* \beta_{\bar b a} \Big) &=&  \delta_{\bar a \bar b} , \label{bogo-restric-11} \\
\sum_a \Big(  \alpha_{\bar a a}^* \beta_{\bar b a}^* -  \beta_{\bar a a}^* \alpha_{\bar b a}^* \Big) &=& 0,
\label{bogo-restric-21z}
\eea
and demand that the Hamiltonian at time $\tau_0$ take the form
\begin{align}
    \hat{\mathcal H}(\tau_0) = \frac{1}{2}\int_{\K} \sum_{\bar a} \omega_{\bar a}(\tau_0) \bigg( \hat b_{\bar a}^\dagger(\K,\tau_0) b_{\bar a}(\K,\tau_0) + \hat b_{\bar a}(\K,\tau_0) b_{\bar a}^\dagger(\K,\tau_0)  \bigg) .
\end{align}

From this perspective too, we can immediately identify the symplectic structure of the Bogoliubov transformation: calling ${\cal T}_c$ the coefficient matrix of~(\ref{multifield-diag-bogo-c}), we see that it obeys Eq.~\eqref{symp}. Further writing it in a block structure, ${\cal T}_c=\mqty(\cal A & \cal B^* \\ \cal B & \cal A^*)$, the constraints~\eqref{bogo-restric-11},~\eqref{bogo-restric-21z} are exactly those imposed by the symplectic group ${\rm Sp}(4,\mathbb{R})$~\cite{Derezinski:2016zhd}.
Using the symplectic structure, one can trivially invert the relation~(\ref{multifield-diag-bogo-c}),
and expand the fields in terms of the adiabatic basis as
\be \label{zeta-b}
\mqty (\hat\zeta_{\K} (\tau_0) \\ \hat\psi_{\K} (\tau_0) \\ \hat\Pi_{\zeta_{\K}} (\tau_0)  \\ \hat\Pi_{\psi_{\K}} (\tau_0) )  = \mqty (\bar\zeta_{\bar1}  & \bar\zeta_{\bar2}  & \bar\zeta_{\bar1}^*  & \bar\zeta_{\bar2}^*  \\ \bar\psi_{\bar1}  & \bar\psi_{\bar2}  & \bar\psi_{\bar1}^*  & \bar\psi_{\bar2}^*  \\ \bar\Pi_{\bar1}^{\zeta}  & \bar\Pi_{\bar2}^{\zeta}   & \bar\Pi_{\bar1}^{\zeta*}  & \bar\Pi_{\bar2}^{\zeta*}    \\ \bar\Pi_{\bar1}^{\psi}   & \bar\Pi_{\bar2}^{\psi}  & \bar\Pi_{\bar1}^{\psi*}  & \bar\Pi_{\bar2}^{\psi*}  ) \mqty (\hat b_{\bar1} ({\K},\tau_0) \\ \hat b_{\bar2} ({\K},\tau_0) \\ \hat b^\dag_{\bar1} (-{\K},\tau_0)  \\ \hat b^\dag_{\bar2} (-{\K},\tau_0) ).
\ee 
The diagonal mode functions are then given by 
\begin{align} \label{bar-zeta-1}
\mqty (\bar\zeta_{\bar1}  & \bar\zeta_{\bar2}  & \bar\zeta_{\bar1}^*  & \bar\zeta_{\bar2}^*  \\ \bar\psi_{\bar1}  & \bar\psi_{\bar2} & \bar\psi_{\bar1}^*  & \bar\psi_{\bar2}^*   \\ \bar\Pi_{\bar1}^{\zeta}   & \bar\Pi_{\bar2}^{\zeta}   & \bar\Pi_{\bar1}^{\zeta*}   & \bar\Pi_{\bar2}^{\zeta*}      \\ \bar\Pi_{\bar1}^{\psi}   & \bar\Pi_{\bar2}^{\psi}   & \bar\Pi_{\bar1}^{\psi*}   & \bar\Pi_{\bar2}^{\psi*}   ) = \mqty(\zeta_1 & \zeta_2 & \zeta_1^* & \zeta_2^* \\ \psi_1 & \psi_2 & \psi_1^* & \psi_2^* \\ \Pi_1^\zeta & \Pi_2^\zeta & \Pi_1^{\zeta *} & \Pi_2^{\zeta *} \\ \Pi_1^\psi & \Pi_2^\psi & \Pi_1^{\psi *} & \Pi_2^{\psi *} ) \mqty( \bar\alpha_{\bar 1 1} & \bar\alpha_{\bar 2 1}  & \bar\beta^*_{\bar 1 1}  & \bar\beta^*_{\bar 2 1}  \\  \bar\alpha_{\bar 1 2}  & \bar\alpha_{\bar 2 2}  & \bar\beta^*_{\bar 1 2}  & \bar\beta^*_{\bar 2 2}  \\ \bar\beta_{\bar 1 1}  & \bar\beta_{\bar 2 1}  & \bar\alpha^*_{\bar 1 1}  & \bar\alpha^*_{\bar 2 1}  \\ \bar\beta_{\bar 1 2}  & \bar\beta_{\bar 2 2}  &  \bar\alpha^*_{\bar 1 2}  & \bar\alpha^*_{\bar 2 2}  ).
\end{align}
By virtue of the Bogoliubov transformation, these satisfy the same generalized Wronskian relations as the original mode functions. (This is also evident from the ${\rm Sp}(4,\mathbb{R})$ structure.)
Finally, from~\eqref{multifield-diag-bogo-c}, the density of eigenstates is given by
\begin{align} \label{n-multi}
    n_{\bar a}(k,\tau_0) = \sum_{a} | \bar\beta_{\bar a a}(\tau_0) |^2 .
\end{align}

As in the single-field case, in order to set up the eigenvalue problem, we consider the Heisenberg equations of motion:
\begin{align}
     \hat\zeta'_\K(\tau_0)   = i \left[ \hat{\mathcal H}(\tau_0), \hat\zeta_\K(\tau_0) \right] &\;,\quad \hat\Pi_{\zeta_\K}(\tau_0)  = i \left[ \hat{\mathcal H}(\tau_0), \hat\Pi_{\zeta_\K}(\tau_0) \right] ,  \\
    \hat\psi'_\K(\tau_0) = i \left[ \hat{\mathcal H}(\tau_0), \hat\psi_\K(\tau_0) \right] &\;,\quad
    \hat\Pi_{\psi_\K}(\tau_0) = i \left[ \hat{\mathcal H}(\tau_0), \hat\Pi_{\psi_\K}(\tau_0) \right] .
\end{align}
Time derivatives in the left-hand side can be written as linear combinations of the fields and momenta as clearly seen in the equations of motion (\ref{eom_b-1}) and (\ref{eom_b-2}). In the right-hand side, we expand the field and Hamiltonian operators in the adiabatic basis and use the following commutators:
\begin{align}
   \left[\hat{\mathcal H}(\tau_0), \hat b_{\bar a}(\K,\tau_0) \right] = - \omega_{\bar a} \hat b_{\bar a}(\K,\tau_0) , \quad \left[ \hat{\mathcal H}(\tau_0), \hat b^\dagger_{\bar a}(-\K,\tau_0) \right] = \omega_{\bar a} \hat b^\dagger_{\bar a}(-\K,\tau_0)  \label{multifield-H-c-commutator}.
\end{align}
Comparing both sides leads to the algebraic relations
\begin{align}
    \mqty(i \omega_{\bar a} & \dfrac{a H \lambda}{\sqrt{2\epsilon}} & \dfrac{1}{2\epsilon a^2} & 0 \\  0 & i \omega_{\bar a} & 0 & \dfrac{1}{a^2}  \\ -2a^2 k^2 \epsilon & 0 & i \omega_{\bar a} & 0  \\ 0 & -a^2\left( k^2+a^2\mu^2 \right) & -\dfrac{a H \lambda}{\sqrt{2\epsilon}} & i \omega_{\bar a}) \mqty( \bar \zeta_{\bar a} \\ \bar \psi_{\bar a} \\ \bar\Pi_{\bar a}^\zeta \\ \bar \Pi_{\bar a}^\psi) = 0 ,
\end{align}
where we have omitted the momentum labels of the modes. Solving for the positive solutions of $\omega$, we obtain
\begin{align}
    \omega_{\bar 1,\bar 2} =  \sqrt{k^2 + \frac{a^2 \mu^2}{2} \mp \frac{a^2 \mu^2}{2}\sqrt{ 1 + 4 \lambda^2 \frac{H^2 k^2}{a^2 \mu^4 }  }}  \label{multifield-eigenfreq-c} .
\end{align}
Note that these are the dispersion relations obtained in Refs.~\cite{Achucarro:2012yr,Castillo:2013sfa} in the Minkowski limit. By employing the adiabatic basis, we see that these arise from multiparticle creation due to the de Sitter expansion and the non-adiabaticity induced by the non-gravitational coupling $\lambda$. 

Having the eigenfrequencies and demanding that the modes respect the Wronskian conditions, we obtain
\begin{align} 
\label{bar-zeta-mode}
    \bar\zeta_{\bar a} = \sqrt{(-1)^{\bar a} \frac{H^2 \lambda^2 \omega_{\bar a}}{4 \epsilon (\omega_{\bar a}^2 - k^2)(\omega_{\bar 2}^2-\omega_{\bar 1}^2)}} e^{i S_{\bar a}} &\;,\quad 
    \bar \Pi_{\bar a}^\zeta = -i \frac{2\epsilon a^2 k^2}{\omega_{\bar a}} \bar\zeta_{\bar a} , \\ 
    \label{bar-psi-mode}
   \bar\psi_{\bar a} = i \frac{\sqrt{2\epsilon}(k^2-\omega_{\bar a}^2)}{a H \lambda \omega_{\bar a}} \bar \zeta_{\bar a} &\;,\quad  
    \bar \Pi_{\bar a}^\psi =  \frac{\sqrt{2\epsilon} a (k^2-\omega_{\bar a}^2)}{H \lambda } \bar \zeta_{\bar a},
\end{align}
where $S_{\bar a}$ is an arbitrary phase.  
In the limit $\lambda \rightarrow 0$, we recover a decoupled system of modes, which are a double copy of the single-field ones~\eqref{adiabatic-mode}: 
\begin{align}
    \omega_{\bar 1} \rightarrow k &\;,\quad  \omega_{\bar 2} \rightarrow \sqrt{k^2+a^2 \mu^2} \,, \label{dcpld-multi-eigenenergy}\\
    \bar\zeta_{\bar 1} \rightarrow \frac{1}{\sqrt{4\epsilon a^2 \omega_{\bar 1}}} e^{i S_{\bar 1}} \; , \; \bar\zeta_{\bar 2} \rightarrow 0 &\;,\quad
    \bar\Pi^{\zeta}_{\bar 1} \rightarrow -i \sqrt{\epsilon a^2 \omega_{\bar 1}} e^{iS_{\bar 1}},  \quad \bar\Pi^\zeta_{\bar 2} \rightarrow 0 \,, \label{dcpld-multi-mode1} \\
    \bar\psi_{\bar 1} \rightarrow 0  \; , \; \bar\psi_{\bar 2} \rightarrow \frac{-i\ \mathrm{sign}(\lambda)}{\sqrt{2 a^2 \omega_{\bar 2}}} e^{i S_{\bar 2}}   &\;,\quad
    \bar\Pi^{\psi}_{\bar 1} \rightarrow 0 \;, \quad \bar\Pi^\psi_{\bar 2} \rightarrow - \mathrm{sign}(\lambda) \sqrt{\frac{ a^2 \omega_{\bar 2}}{2}} e^{iS_{\bar 2}}  \,. \label{dcpld-multi-mode2}
\end{align}

Once we have dealt with the eigenvalue problem, we can compute the Bogoliubov coefficients $( \bar \alpha_{\bar a a}, \bar \beta_{\bar a a})$ of the transformation~\eqref{multifield-diag-bogo-c}, which brings us from a state with mode functions $(\zeta,\psi)$ to the one corresponding to adiabatic modes $(\bar\zeta,\bar\psi)$. These are obtained by 
inverting Eq.~\eqref{bar-zeta-1}, to obtain 
\begin{align} \label{bar-aleph}
  \bar \alpha_{\bar a a} &= -i \left( \bar \zeta_{\bar a} \Pi_a^{\zeta *} + \bar \psi_{\bar a} \Pi_a^{\psi *} - \bar \Pi^\zeta_{\bar a} \zeta^*_a - \bar \Pi^\psi_{\bar a} \psi_a^*  \right) , \\ \label{bar-beta}
    \bar\beta_{\bar a a} &= i \left( \bar \zeta_{\bar a} \Pi_a^\zeta + \bar \psi_{\bar a} \Pi_a^\psi - \bar \Pi^\zeta_{\bar a} \zeta_a - \bar \Pi^\psi_{\bar a} \psi_a  \right).
\end{align}
With the mode functions~\eqref{bar-zeta-mode} and the Bogoliubov coefficients $\bar\beta_{\bar a a}$ at hand, we can now compute the eigenstate number-density from Eq.~\eqref{n-multi}:
\begin{align} \label{n_a}
    n_{\bar a} = \frac{|\bar \zeta_{\bar a}|^2}{a^2 H^2 \lambda^2 \omega_{\bar a}^2} \sum_a \Big| a H \lambda (\omega_{\bar a} \Pi_a^\zeta + i 2\epsilon a^2 k^2 \zeta_a) + i \sqrt{2\epsilon}(k^2-\omega_{\bar a}^2) (\Pi_a^\psi + i \omega_{\bar a}\psi_a) \Big|^2.
\end{align}
This expression is the main result of the present work. It yields the number-density of Hamiltonian eigenstates of a state $|\Omega_1\rangle\times|\Omega_2\rangle$, satisfying $ \hat a_1  \hat a_2 |\Omega_1\rangle\times|\Omega_2\rangle =0$, with corresponding mode functions $\zeta_a,\psi_a$. Note that for $\lambda=0$, that is, for a straight trajectory, the eigenstate number-densities reduce to
\begin{align}
    n_{\bar 1} &= \frac{2\epsilon a^2}{2 k} \sum_a \Big( |\zeta_a'|^2 + k^2 |\zeta_a|^2 \Big) - \frac{1}{2} , \label{numb-density-dcpld-1}\\
    n_{\bar 2} &= \frac{a^2}{2 \sqrt{k^2+a^2 \mu^2}} \sum_a \Big( |\psi_a'|^2 + \left(k^2+a^2\mu^2 \right) |\psi_a|^2 \Big) - \frac{1}{2} , \label{numb-density-dcpld-2}
\end{align}
which, as expected, coincide with the single-field result~(\ref{number-density-bog}) for each degree of freedom. 

Finally, as in the single-field case, we may write the exact solution of the coupled system~\eqref{eom_b-1},~\eqref{eom_b-2} as a canonical transformation over the adiabatic initial conditions. This is simply done by inverting Eq.~\eqref{bar-zeta-1},
with the adiabatic modes given by Eqs.~(\ref{bar-zeta-mode},~\ref{bar-psi-mode}). 
These Bogoliubov coefficients can be obtained as solutions to a first-order system analogous to~\eqref{diag-eom}. Setting the phases $S_{\bar a}'=-\omega_{\bar a}$ yields an equation of the form
\begin{align}
    \mqty(\bar\alpha_{\bar a a} \\ \bar\beta_{\bar a a})' = \sum_{\bar b} \mathcal M_{\bar b}^{(\mathrm{ad})}(k,\tau) \mqty(\bar\alpha_{\bar b a} \\ \bar\beta_{\bar b a}),
\end{align}
where each term in each matrix element of $\mathcal M^{(\mathrm{ad})}$ is proportional to one of the following combinations:
\begin{align}
    a H , \ \frac{\lambda'}{\lambda}, \ \frac{\omega_{\bar a}'}{\omega_{\bar a}}, \ \ \frac{\omega_{\bar 1}'}{\omega_{\bar 2}} - \frac{\omega_{\bar 2}'}{\omega_{\bar 1}},
\end{align}
which measure the adiabaticity of the system. Solutions with constant coefficients correspond to the lowest-order multifield adiabatic approximation, which is valid in the adiabatic regime or for a small duration of non-adiabatic dynamics. Remarkably, the decomposition, with some minor corrections, remains valid for imaginary $\omega_{\bar a}$, leading to a dominant exponential growth of the modes. These are the instabilities that are responsible for primordial black hole formation in this context~\cite{Palma:2020ejf,Fumagalli:2020adf}, which constitute a general feature of rapid turn inflationary models. It would be interesting to study this case using the Stokes phenomenon~\cite{Hashiba:2021npn,Chung:2024lky} or other methods~\cite{Jenks:2024fiu} like \emph{e.g.} steepest descent~\cite{Chung:1998bt}.

Finally, let us note that multifield models with turning trajectories have been analyzed in~\cite{Christodoulidis:2023eiw}, where the authors point out the role of Hubble friction when one uses WKB approximations to de Sitter dynamics~\cite{Bjorkmo:2019qno}. In our language, their conclusion is reflected in the fact that the Bogoliubov coefficients $\left(\alpha_{\bar a a},\beta_{\bar a a}\right)$ of the adiabatic basis are time-dependent, with the WKB (constant) value attained only as a first approximation.
%%%%
%%%%

\subsection{The Bunch-Davies basis}
\label{time-dependent-Bog-multi-field}

In Sec.~\ref{time-dependent-Bog-single-field}, we saw how the interaction terms emerging from slow-roll parameters $\epsilon$ and $\eta$ can be captured by a specific time-dependent Bogoliubov transformation. Here we will generalize that discussion to the multifield case and derive the solution of the coupled system as a canonical transformation on BD initial conditions. Our focus here is not a temporal deformation of the de Sitter geometry as in the single-field case but a temporal deformation of the dynamics due to the mixing of fluctuations.

The use of this Hilbert-space basis instead of the adiabatic one of Sec.~\ref{Sec:mf-adiab}, will allow us to study particle production from emergent excited states~\cite{Fumagalli:2020nvq,Fumagalli:2021mpc} produced via non-gravitational couplings. We thus want to impose initial conditions not at the Minkowski limit but at some finite subhorizon scale. For example, in this way, one can study setups where a sudden turn takes place at some definite moment $\tau_\lambda > -\infty$. In these scenarios of emergent excited states, the modes decouple for early times $\tau<\tau_\lambda$, however, their mode functions are given by the BD form (or the quasi-de Sitter form) instead of the Minkowski plane-wave (the adiabatic modes). Moreover, in such models it is the particle number-density in this basis that plays an important role in backreaction computations~\cite{Holman:2007na}.

To proceed, let us first consider the solutions of Eqs.~(\ref{eom_b-1},~\ref{eom_b-2}) in the particular case $\lambda = 0$ (that is, in the absence of multifield interactions). These solutions have the form:
\bea
\zeta_{1} (k , \tau) = \zeta_0 (k , \tau) , \qquad \zeta_{2} (k , \tau) = 0 , \qquad \psi_{1} (k , \tau) = 0 , \qquad \psi_{2} (k , \tau) = \psi_0 (k , \tau) ,  \label{non-interacting-modes}
\eea
where $ \zeta_0 (k , \tau)$ and $\psi_0 (k , \tau) $ are single-field solutions of the following system of decoupled equations of motion:
\bea
\frac{\de}{\de\tau} \Pi_0^\zeta  + 2 \epsilon a^2 k^2   \zeta_0 &=&0  , \qquad  \Pi_0^\zeta = 2 \epsilon a^2 \frac{\de}{\de\tau} \zeta_0  , \label{single-zeta-psi-1}  \\
\frac{\de}{\de\tau} \Pi_0^\psi + a^2 \left( k^2  + a^2 \mu^2 \right)  \psi_0  &=& 0 , \qquad  \Pi_0^\psi =  a^2 \frac{\de}{\de\tau} \psi_0 .  \label{single-zeta-psi-2} 
\eea
Having these single-field solutions at hand, we can now consider a time-dependent Bogoliubov transformation defining new time-dependent operators  $\hat c^\dag_\zeta  (\K, \tau)$, $\hat c_\zeta  (\K, \tau)$,  $\hat c_\psi^\dag (\K, \tau)$ and $\hat c_\psi (\K, \tau)$, in terms of $\hat a^\dag_a  (\K)$ and $\hat a_a  (\K)$. Explicitly, the transformation reads
\be
\mqty (\hat c_\zeta (\K, \tau)  \\ \hat c_\psi (\K, \tau) \\ \hat{c}^\dag_\zeta (-\K, \tau)  \\ \hat{c}^\dag_\psi (-\K, \tau) )    = \mqty ( \alpha_{\zeta 1} & \alpha_{\zeta 2} & \beta^*_{\zeta 1} & \beta^*_{\zeta 2} \\ \alpha_{\psi 1} & \alpha_{\psi 2} & \beta^*_{\psi 1} & \beta^*_{\psi 2} \\ \beta_{\zeta 1} & \beta_{\zeta 2} & \alpha^*_{\zeta 1} & \alpha^*_{\zeta 2} \\ \beta_{\psi 1} & \beta_{\psi 2} & \alpha^*_{\psi 1} & \alpha^*_{\psi 2})  \mqty (\hat a_1 (\K, \tau)  \\ \hat a_2 (\K, \tau) \\ \hat{a}^\dag_1 (-\K, \tau)  \\ \hat{a}^\dag_2 (-\K, \tau) ). \label{multi-bogo-transf-1} 
\ee 
Again, the symplectic structure ensures the canonical character of the Bogoliubov transformation, that is, after imposing
\bea
\sum_a \Big(  \alpha_{I a} \alpha_{J a}^* -  \beta_{I a}^* \beta_{J a} \Big) &=&  \delta_{IJ} , \label{bogo-restric-1} \\
\sum_a \Big(  \alpha_{I a} \beta_{J a}^* -  \beta_{I a}^* \alpha_{J a} \Big) &=& 0,
\label{bogo-restric-2}
\eea
the new operators satisfy $\big[ c_{I} (\K) , c^{\dag}_{J} (\K') \big] = (2 \pi)^3 \delta_{IJ} \delta^{(3)} (\K - \K')$, with $I, J \in (\zeta, \psi)$.

Just as we did in the single-field case, we may choose a specific transformation 
by imposing a condition analogous to that leading to Eq.~(\ref{Q-alpha-beta}): we demand that the full fields $\zeta (\tau, {\K})$ and $\psi (\tau, {\K})$ (and their momenta), satisfying the coupled system of equations, take the form 
\be
\mqty( \hat \zeta (\tau, {\K}) \\ \hat \psi (\tau, {\K})  \\  \hat \Pi_\zeta (\tau, {\K}) \\ \hat \Pi_\psi (\tau, {\K}) ) =  \mqty(  \zeta_0 & 0 & \zeta_0^* & 0 \\ 0 & \psi_0 & 0 & \psi_0^* \\ \Pi_0 ^{\zeta} & 0 & \Pi_0 ^{\zeta *} & 0 \\ 0 & \Pi_0 ^{\psi} & 0 & \Pi_0 ^{\psi*} ) \mqty (\hat c_\zeta (\K, \tau)  \\ \hat c_\psi (\K, \tau) \\ \hat{c}^\dag_\zeta (-\K, \tau)  \\ \hat{c}^\dag_\psi (-\K, \tau) ) .  \label{multi-Fourier-5}  
\ee
In other words, we impose that the field operators take the form of their single-field counterparts in the $\hat c$-basis. Upon equating (\ref{multi-Fourier}) to (\ref{multi-Fourier-5}) and using~\eqref{multi-bogo-transf-1}, the relation between the $\hat a$-basis and the $\hat c$-basis mode functions becomes
\be
\mqty (\zeta_1 & \zeta_2  & \zeta_1^* & \zeta_2^*  \\ \psi_1  & \psi_2  & \psi_1^*  & \psi_2^*   \\ \Pi_1^{\zeta}   & \Pi_2^{\zeta}  & \Pi_1^{\zeta*}   & \Pi_2^{\zeta*}  \\ \Pi_1^{\psi}  & \Pi_2^{\psi}  & \Pi_1^{\psi*} & \Pi_2^{\psi*} ) = \mqty(  \zeta_0 & 0 & \zeta_0^* & 0 \\ 0 & \psi_0 & 0 & \psi_0^* \\ \Pi_0 ^{\zeta} & 0 & \Pi_0 ^{\zeta *} & 0 \\ 0 & \Pi_0 ^{\psi} & 0 & \Pi_0 ^{\psi*} ) \mqty ( \alpha_{\zeta 1} & \alpha_{\zeta 2} & \beta^*_{\zeta 1} & \beta^*_{\zeta 2} \\ \alpha_{\psi 1} & \alpha_{\psi 2} & \beta^*_{\psi 1} & \beta^*_{\psi 2} \\ \beta_{\zeta 1} & \beta_{\zeta 2} & \alpha^*_{\zeta 1} & \alpha^*_{\zeta 2} \\ \beta_{\psi 1} & \beta_{\psi 2} & \alpha^*_{\psi 1} & \alpha^*_{\psi 2})   .   \label{multi-bog-1}
\ee

Now, notice that the conjugate momenta of the fields $\zeta_a,\psi_a$ are given by $ \Pi^\zeta_a = 2 \epsilon a^2 D_\tau \zeta_a$ and $\Pi^\psi_a =  a^2  \psi_a'$, whereas the ones of the free fields read $\Pi_0^\zeta = 2 \epsilon a^2  \zeta_0'$ and $\Pi_0^\psi =  a^2 \psi'$. As a consequence, the Bogoliubov coefficients must satisfy
\bea
\zeta_0 \frac{\de}{\de\tau} \alpha_{\zeta a} +  \zeta_0^* \frac{\de}{\de\tau} \beta_{\zeta a} &=& a H \frac{\lambda}{\sqrt{2 \epsilon}} \Big( \psi_0 \alpha_{\psi a} +  \psi^*_0 \beta_{\psi a} \Big) , \label{pre-bog-1} \\
\psi_0  \frac{\de}{\de\tau} \alpha_{\psi a} + \psi_0^* \frac{\de}{\de\tau} \beta_{\psi a} &=& 0 . \label{pre-bog-2}
\eea
In addition, the mode functions $ \zeta_a$ and $ \psi_a$ satisfy the coupled system of equations~(\ref{eom_b-1},~\ref{eom_b-2}), whereas $ \zeta^0_a$ and $ \psi^0_a$ respect the single-field equations~(\ref{single-zeta-psi-1},~\ref{single-zeta-psi-2}). As a consequence, we obtain additional equations satisfied by the Bogoliubov coefficients:
\bea
 \Pi_0 ^{\zeta}  \frac{\de}{\de\tau} \alpha_{\zeta a} +  \Pi_0 ^{\zeta *}  \frac{\de}{\de\tau} \beta_{\zeta a}  &=& 0 , \label{pre-bog-3} \\
 \Pi_0 ^{\psi}  \frac{\de}{\de\tau} \alpha_{\psi a} +  \Pi_0 ^{\psi *}  \frac{\de}{\de\tau} \beta_{\psi a}  &=& -   a H \frac{\lambda}{\sqrt{2 \epsilon} }  \Big(  \alpha_{\zeta a}  \Pi_0 ^{\zeta}  +  \beta_{\zeta a}  \Pi_0 ^{\zeta *}  \Big) . \label{pre-bog-4}
\eea
One can now put together Eqs.~(\ref{pre-bog-1})-(\ref{pre-bog-4}) to obtain a set of first-order equations dictating the evolution of the Bogoliubov coefficients. One finds:
\be
\frac{1}{k} \frac{\de}{\de\tau} \left(\begin{array}{c} \alpha_{\zeta a} \\   \beta_{\zeta a} \\   \alpha_{\psi a} \\  \beta_{\psi a} \end{array}\right)  = \mathcal M (k \tau)  \left(\begin{array}{c} \alpha_{\zeta a} \\   \beta_{\zeta a} \\   \alpha_{\psi a} \\  \beta_{\psi a} \end{array}\right) , \qquad \mathcal M  (k \tau) =  \left(\begin{array}{cc} 0 & \mathcal P  (k \tau) \\  \mathcal Q  (k \tau) & 0   \end{array}\right) , \label{eq-alpha-beta-multi-1}
\ee
where $\mathcal P (k \tau)$ and $\mathcal Q (k \tau)$ are $2 \times 2$ matrices given by
\bea
\mathcal P (k \tau) &=&  - i \frac{a H}{k} \frac{\lambda}{\sqrt{2 \epsilon}} \left(\begin{array}{cc} \psi_0 \Pi_0^{\zeta *}     &   \psi_0^* \Pi_0^{\zeta *} \\   - \psi_0 
 \Pi_0^{\zeta}  & - \psi_0^* \Pi_0^{\zeta}   \end{array}\right)  , \\
\mathcal Q (k \tau) &=& - i \frac{a H}{k} \frac{\lambda}{\sqrt{2 \epsilon}} \left(\begin{array}{cc}  \psi_0^* \Pi_0 ^{\zeta}  &  \psi_0^* \Pi_0 ^{\zeta *}    \\  - \psi_0 \Pi_0^{\zeta}  & - \psi_0 \Pi_0^{\zeta *}  \end{array}\right).  
\eea

As already mentioned, the initial conditions respected by the Bogoliubov coefficients may be set by requiring that at some point (which could also be the infinite past) the solutions $\zeta_a$ and $\psi_a$ coincide with the non-interacting fields as in Eq.~(\ref{non-interacting-modes}). This would require setting $( \alpha_{\zeta 1}  ,\beta_{\zeta 1} , \alpha_{\psi_1} ,  \beta_{\psi 1} ) =  (1 , 0,0,0)$ and $( \alpha_{\zeta 2}  ,\beta_{\zeta 2} , \alpha_{\psi_2} ,  \beta_{\psi 2} ) =  (0 , 0,1,0)$ at some time $\tau_{\rm ini}$ such that $k | \tau_{\rm ini} | \gg 1$. With this initial condition, the formal solution to Eq.~(\ref{eq-alpha-beta-multi-1}) can be written as
\be
\left(\begin{array}{c} \alpha_{\zeta a} \\   \beta_{\zeta a} \\   \alpha_{\psi a} \\  \beta_{\psi a} \end{array}\right)   = \mathcal T \exp \left\{ \int^{k \tau}_{k \tau_{\rm ini} } \!\!\!  \de x \, \mathcal M  (x) \right\} \left(\begin{array}{c} \alpha_{\zeta a} \\   \beta_{\zeta a} \\   \alpha_{\psi a} \\  \beta_{\psi a} \end{array}\right)_{\rm ini}  . \label{sol-Dyson-2}
\ee

To summarize the present discussion, let us stress the main point of this section: The introduction of time-dependent Bogoliubov coefficients allowed us to split the multifield system of equations (\ref{eom_b-1},~\ref{eom_b-2}) into two groups. The first set corresponds to the equations of motion (\ref{single-zeta-psi-1},~\ref{single-zeta-psi-2}) describing the evolution of two decoupled fields $\zeta_0$ and $\psi_0$. These modes only experience the Hubble flow  of the background encoded in the scale factor $a(\tau)$. Once the solutions $\zeta_0$ and $\psi_0$ are known, one may study the effects of the multifield interaction $\lambda$ via the first-order equation of motion (\ref{eq-alpha-beta-multi-1}). An additional feature of this formalism is that it allows us to gain a clear picture of the particle states generated exclusively from the coupling $\lambda$ thanks to the standard language of Bogoliubov transformations. Gravitational (\emph{e.g.} slow-roll) corrections may be incorporated independently from non-gravitational effects (non-geodesic motion). In other words, $\zeta_0$ and $\psi_0$ can either be the BD mode functions or the quasi-de Sitter mode functions\footnote{See also~\cite{Bianchi:2024jmn} for a discussion on disentangling features coming from the Hubble-flow and from non-gravitational effects like the sound-speed in single-field inflation.} of Seq.~\ref{subsec:ns}, computed to any order in slow roll.

\subsection{Massless isocurvature fields}

A particularly relevant case that is worth examining in more detail, corresponds to the situation where the entropy mass remains zero: $\mu = 0$. In this instance, the single-field mode functions assume their standard BD form:
\bea
\zeta_0 (k, \tau) &=&  \frac{ i H }{2\sqrt{\epsilon k^3}} \left( 1 +   i k \tau \right) e^{- i k \tau } , \label{Z-0} \\
\Pi^\zeta_0 (k ,\tau) &=&  \frac{ i   \sqrt{ \epsilon k} }{ H \tau}   e^{- i k \tau },\label{Pi-Z-0} \\
\psi_0 (k, \tau) &=& \frac{i H}{\sqrt{2k^3}} \left( 1 + i k \tau  \right) e^{- i k \tau } , \label{Psi-less-0}  \\
\Pi^\psi_0 (k, \tau) &=&    \frac{ i }{ H \tau} \sqrt{\frac{k}{2}}  e^{- i k \tau } . \label{Pi-Psi-less-0}
\eea
One can readily verify that in this case $\psi_0$ does not decay (as encountered in the massive case $\mu \neq 0$) and the effects due to $\lambda$ between the fields become maximal during and after horizon crossing. With these expressions the matrices $\mathcal P$ and $\mathcal Q$ become
\bea
\mathcal P (x) &=&   \frac{i \lambda}{2 x^2}  \left(\begin{array}{cc}    1 + i x     & -  \left( 1 - i x  \right) e^{2 i x } \\       \left( 1 + i x  \right) e^{- 2 i x }  & -   \left( 1 - i x \right)  \end{array}\right) ,  \label{massless-P} \\
\mathcal Q (x) &=&  \frac{i \lambda}{2 x^2}  \left(\begin{array}{cc}    1 - i x  &    -  \left( 1 - i x \right) e^{ 2 i x }   \\     \left( 1 + i x \right) e^{- 2 i x } & -  \left( 1 + i x \right) \end{array}\right).    \label{massless-Q}
\eea
We will examine analytical solutions of Eq.~(\ref{eq-alpha-beta-multi-1}) for the particular case of constant $\lambda$ in Sec.~\ref{sec:Bog-constant-lambda}. 

In the study of inflation, we are typically interested in the fields on long wavelengths $k |\tau | \ll 1$. In this limit, Eq.~(\ref{multi-bog-1}) yields
\bea
\zeta_a &=&  \frac{ i H }{2\sqrt{\epsilon k^3}} ( \alpha_{\zeta a} -  \beta_{\zeta a}  ) , \label{asympt-fields-1}  \\
\psi_a &=&  \frac{ i H }{\sqrt{ 2 k^3}} ( \alpha_{\psi a} -  \beta_{\psi a}  ) ,  \label{asympt-fields-2}
\eea
where the Bogoliubov coefficients are to be evaluated on long wavelengths. Similar expressions may be easily derived for the conjugate momenta. 
The power spectra of $\zeta$ and $\psi$ are then found to be
\bea
P_{\zeta} &=&  \frac{ H^2 }{ 4 \epsilon k^3} \Big(  | \alpha_{\zeta 1} -  \beta_{\zeta 1} |^2 +  | \alpha_{\zeta 2} -  \beta_{\zeta 2} |^2 \Big) ,  \label{power-zeta-alpha} \\
P_{\psi} &=&  \frac{ H^2 }{ 2 k^3} \Big(  | \alpha_{\psi 1} -  \beta_{\psi 1} |^2 +  | \alpha_{\psi 2} -  \beta_{\psi 2} |^2 \Big) .  \label{power-psi-alpha}
\eea
In addition, the cross power spectrum between $\zeta$ and $\psi$ reads
\be
P_{\zeta \psi} =   \frac{ H^2 }{2\sqrt{2 \epsilon } k^3} \Big(  ( \alpha_{\zeta 1} -  \beta_{\zeta 1}  )  ( \alpha^*_{\psi 1} -  \beta^*_{\psi 1}  ) + ( \alpha_{\zeta 2} -  \beta_{\zeta 2}  )  ( \alpha^*_{\psi 2} -  \beta^*_{\psi 2}  )  \Big) , \label{power-zeta-psi-alpha}
\ee
which may be shown to be real thanks to the symplectic restrictions~(\ref{bogo-restric-1},~\ref{bogo-restric-2}).

\setcounter{equation}{0}
\section{Bogoliubov coefficients for constant kinetic mixing} \label{sec:Bog-constant-lambda}

As an application of the previous results, let us examine the particular case whereby $\mu=0$ and $\lambda$ is small and constant. Here too, we will neglect the evolution due to the Hubble flow, measured by the slow-roll parameters $\epsilon$ and $\eta$ discussed in the single-field case. As already mentioned, when necessary, these corrections can be incorporated in the calculation in a straightforward manner. Before explicitly computing the analytical solutions of Eq.~(\ref{eq-alpha-beta-multi-1}) valid for small $\lambda$, we examine the asymptotic form of the solutions in the short- and long-wavelength regimes valid for any value of $\lambda$.

\subsection{Asymptotic behaviour for constant, arbitrary $\lambda$}

To start with, let us analyze the differential equations~(\ref{eq-alpha-beta-multi-1}) in the short-wavelength limit, $k |\tau | \gg 1$. Here, one finds that the matrices $\mathcal P$ and $\mathcal Q$ acquire the form
\bea
\mathcal P (x) =  - \frac{ \lambda}{2 x}  \left(\begin{array}{cc}   1  &  0 \\     0  &  1  \end{array}\right) , \qquad \mathcal Q (x) =  \frac{ \lambda}{2 x}  \left(\begin{array}{cc}   1  &    0  \\   0 &  1   \end{array}\right).  
\eea
Then (\ref{eq-alpha-beta-multi-1}) reduces to the following set of equations:
\bea
x \frac{\dd  \alpha_{\zeta a}}{\dd x} &=& - \frac{\lambda}{2} \alpha_{\psi a} , \qquad x \frac{\dd  \alpha_{\psi a}}{\dd x} =  \frac{\lambda}{2} \alpha_{\zeta a} , \\
x \frac{\dd  \beta_{\zeta a}}{\dd x} &=& -  \frac{\lambda}{2} \beta_{\psi a} , \qquad x \frac{\dd  \beta_{\psi a}}{\dd x} =   \frac{\lambda}{2} \beta_{\zeta a} . 
\eea
The solutions are straightforward to obtain. Imposing the initial conditions $( \alpha_{\zeta 1}  ,\beta_{\zeta 1} , \alpha_{\psi_1} ,  \beta_{\psi 1} ) =  (1 , 0,0,0)$ and $( \alpha_{\zeta 2}  ,\beta_{\zeta 2} , \alpha_{\psi_2} ,  \beta_{\psi 2} ) =  (0 , 0,1,0)$ at initial time $\tau_{\rm ini}$ (such that $k | \tau_{\rm ini} | \gg 1$), we find
\bea
\alpha_{\zeta 1} ( k \tau) = \cos \left( \frac{\lambda}{2} \ln (\tau / \tau_{\rm ini} ) \right) , \qquad \alpha_{\psi 1} (k \tau) = \sin \left( \frac{\lambda}{2} \ln (\tau / \tau_{\rm ini} ) \right)  , \\
\alpha_{\zeta 2} ( k \tau ) = - \sin \left( \frac{\lambda}{2} \ln (\tau / \tau_{\rm ini} ) \right) , \qquad \alpha_{\psi 2} ( k \tau) = \cos \left( \frac{\lambda}{2} \ln (\tau / \tau_{\rm ini} ) \right)  , 
\eea
with $\beta_{\zeta 1} = \beta_{\psi 1} = \beta_{\zeta 2} = \beta_{\psi 2} = 0$. These expressions reveal how the coupling $\lambda$ alters the oscillatory behavior of the the Bunch-Davies mode functions at small scales. It should be clear that $\tau_{\rm ini}$ appears as a phase in the mode functions but it cannot appear in any observables. We will ratify this expectation in a moment, when we explicitly compute the field-spectra.  

On the other hand, in the long-wavelength limit, $k |\tau| \ll 1$, the matrices $\mathcal P$ and $\mathcal Q$ become
\bea
\mathcal P (x) &=&   \frac{i \lambda}{2 x^2}  \left(\begin{array}{cc}    1 + i x    \, & -  ( 1 +  i x ) \\       1 -  i x \, & -   ( 1 - i x  )  \end{array}\right) , \\
\mathcal Q (x) &=&  \frac{i \lambda}{2 x^2}  \left(\begin{array}{cc}   1 - i x \, &    -  ( 1  +  i x )   \\     1 -  i x  \,& -  ( 1 + i x ) \end{array}\right),  
\eea
in which case, Eq.~(\ref{eq-alpha-beta-multi-1}) can also be easily solved:
\bea
\alpha_{\psi a} - \beta_{\psi a} &=& C_{a1}, \label{asympt-sol-alpha-beta-1} \\
\alpha_{\zeta a} -   \beta_{\zeta a}  &=& -  \lambda C_{a1} \ln k |\tau| + C_{a2} , \\
 \alpha_{\zeta a} +   \beta_{\zeta a}  &=& -   \frac{i}{k\tau} \lambda  C_{a1} + C_{a3} , \\
\alpha_{\psi a} + \beta_{\psi a}   &=&   \frac{i}{k \tau} \lambda^2   C_{a1} \left( 2 + \ln  k |\tau|  \right)  -   \frac{i}{k \tau} \lambda  C_{a2}   +   \lambda C_{a3}  \ln k|\tau|  + C_{a4} , \label{asympt-sol-alpha-beta-4}
\eea
where $C_{a1}$, $C_{a2}$, $C_{a3}$ and $C_{a4}$ are integration constants. 
These depend not only on the coupling $\lambda$ but also on the initial conditions imposed at $\tau_{\rm ini}$, while they are further restricted by the conditions (\ref{bogo-restric-1},~\ref{bogo-restric-2}). 

Moving back to Eqs.~(\ref{asympt-fields-1},~\ref{asympt-fields-2}) we can now write down the asymptotic form of the fields in this regime. We obtain:
\bea
\zeta_a &=&  \frac{ i H }{2\sqrt{\epsilon k^3}} ( C_{a2}   -  \lambda C_{a1} \ln k |\tau| ) ,  \label{zeta-asym} \\
\psi_a &=&  \frac{ i H }{\sqrt{ 2 k^3}} C_{a1} , 
\eea
while similar expressions hold for the conjugate momenta.
We emphasize that these solutions are valid for any value of $\lambda$. From Eqs.~\eqref{power-zeta-alpha}-\eqref{power-zeta-psi-alpha}, the power spectra are then given by 
\bea
P_{\zeta} &=&  \frac{ H^2 }{ 4 \epsilon k^3} \Big(  \Big| \lambda C_{11} \ln k |\tau| - C_{12} \Big|^2 +  \Big| \lambda C_{21} \ln k |\tau| - C_{22} \Big|^2 \Big) ,  \\
P_{\psi} &=&  \frac{ H^2 }{ 2 k^3} \Big(  | C_{11}|^2 +  |C_{21}|^2 \Big) , \\
P_{\zeta \psi} &=&   \frac{ H^2 }{2\sqrt{2 \epsilon } k^3} \Big(  -  \lambda |C_{11}|^2 \ln k |\tau| + C_{12}C_{11}^*    -  \lambda |C_{21}|^2 \ln k |\tau| + C_{22}  C_{21}^*    \Big) .
\eea
Note that in the expression~\eqref{zeta-asym}, $H$ and $\epsilon$ are constants, that is, we are considering exact de Sitter dynamics. Comparing this result with the asymptotic from~\eqref{zeta-freeze} of $\zeta$ in the case of a quasi-de Sitter expansion, we see that there is no way for slow-roll corrections to mimic the effect of the non-gravitational coupling, which is reflected in the fact that, here, the curvature perturbation does not freeze on superhorizon scales.

\subsection{Solution for small $\lambda$}

We now move on to derive solutions for the Bogoliubov coefficients valid up to second order in $\lambda$ for the massless case $\mu = 0$. To do so, we may insert the matrices $\mathcal P$ and $\mathcal Q$ of Eqs.~(\ref{massless-P}) and (\ref{massless-Q}) back into Eq.~(\ref{sol-Dyson-2}), expand up to second order in $\lambda$, and perform the integrals over the physical momentum. This can be done exactly~\cite{Parra} but let us present here the solutions in the long-wavelength limit, $k | \tau | \ll 1$. The first-mode Bogoliubov coefficients are given by
\bea \label{multibogoz-1}
\alpha_{\zeta 1} &=& 1 - \frac{\lambda^2}{4}\bigg( \frac{i \pi}{2} - w - \ln (-k\tau_{\rm ini}) \bigg) \bigg( \frac{i}{k\tau} + \ln(-k\tau) \bigg) \nn \\
&& \hspace{0.2cm} - \frac{\lambda^2}{8}\bigg( \big[\ln(-k\tau_{\rm ini})\big]^2 - w^2 + i \pi w - \frac{\pi^2}{4} \bigg)  ,\\
\beta_{\zeta 1} &=&  \frac{\lambda^2}{4} \bigg(\frac{i \pi}{2}-w-\ln(-k \tau_{\mathrm{ini}}) \bigg) \bigg( \frac{-i}{k\tau} + \ln(-k\tau) + \ln(-k\tau_{\rm ini}) \bigg) \nn \\
& & + \frac{\lambda^2}{4}\bigg( 2 + \frac{\pi^2}{12} + \big[\ln(-k\tau_{\rm ini})\big]^2 - w^2 + i \pi w \bigg) ,\\  
\alpha_{\psi 1} &=&  - \frac{ \lambda}{2}    \left(   \frac{i}{k \tau}  - \ln (\tau / \tau_{\rm ini})  \right)  , \\
\beta_{\psi 1} &=& -\frac{\lambda}{2}\bigg( \frac{i}{k\tau} + \frac{i \pi}{2} - w - \ln(-k\tau)  \bigg)  , 
\eea
where $w \equiv   \alpha_{\rm E} - 2 + \ln 2 $, with $ \alpha_{\rm E}$ the Euler-Mascheroni constant. On the other hand, the integration of the second-mode coefficients yields:
\bea
\alpha_{\zeta 2} &=& -  \frac{ \lambda}{2}    \left(  \frac{i}{k \tau}  + \ln (\tau / \tau_{\rm ini})  \right)  , \\
\beta_{\zeta 2}  &=& -\frac{\lambda}{2}\bigg( \frac{i}{k\tau} + \frac{i \pi}{2} - w - \ln(-k\tau)  \bigg)  , \\
\alpha_{\psi 2} &=& 1 - \frac{\lambda^2}{4} \bigg(\frac{i \pi}{2} - w - \ln (-k\tau_{\mathrm{ini}})\bigg) \bigg( \frac{i}{k\tau} + \ln(-k\tau) \bigg) + \frac{i \lambda^2}{k\tau}\bigg(1 + \frac{1}{2}\ln(\tau/\tau_{\rm ini}) \bigg) \nn  \\
&& \hspace{0.2cm} - \frac{\lambda^2}{8}\bigg(\big[\ln(-k\tau_{\rm ini}) \big]^2 - w^2 + i \pi w - \frac{\pi^2}{4}\bigg) , \\
\beta_{\psi 2} &=&  -\frac{\lambda^2}{4} \bigg(\frac{i \pi}{2}-w-\ln(-k \tau_{\mathrm{ini}}) \bigg) \bigg( \frac{i}{k\tau} + \ln(-k\tau) + \ln(-k\tau_{\rm ini}) \bigg) \nn \\
& & + \frac{\lambda^2}{4}\bigg( 2 - \frac{\pi^2}{12} - \big[\ln(-k\tau_{\rm ini})\big]^2 + w^2 - i \pi w \bigg) + \frac{i \lambda^2}{k\tau}\bigg(1 + \frac{1}{2}\ln(\tau/\tau_{\rm ini}) \bigg).   \label{multibogoz-2}
\eea

%plot
\begin{figure}[h!]
\centering
\begin{subfigure}{0.49\textwidth}
    \hspace*{-1.5cm}
    \includegraphics[width=1.2\linewidth]{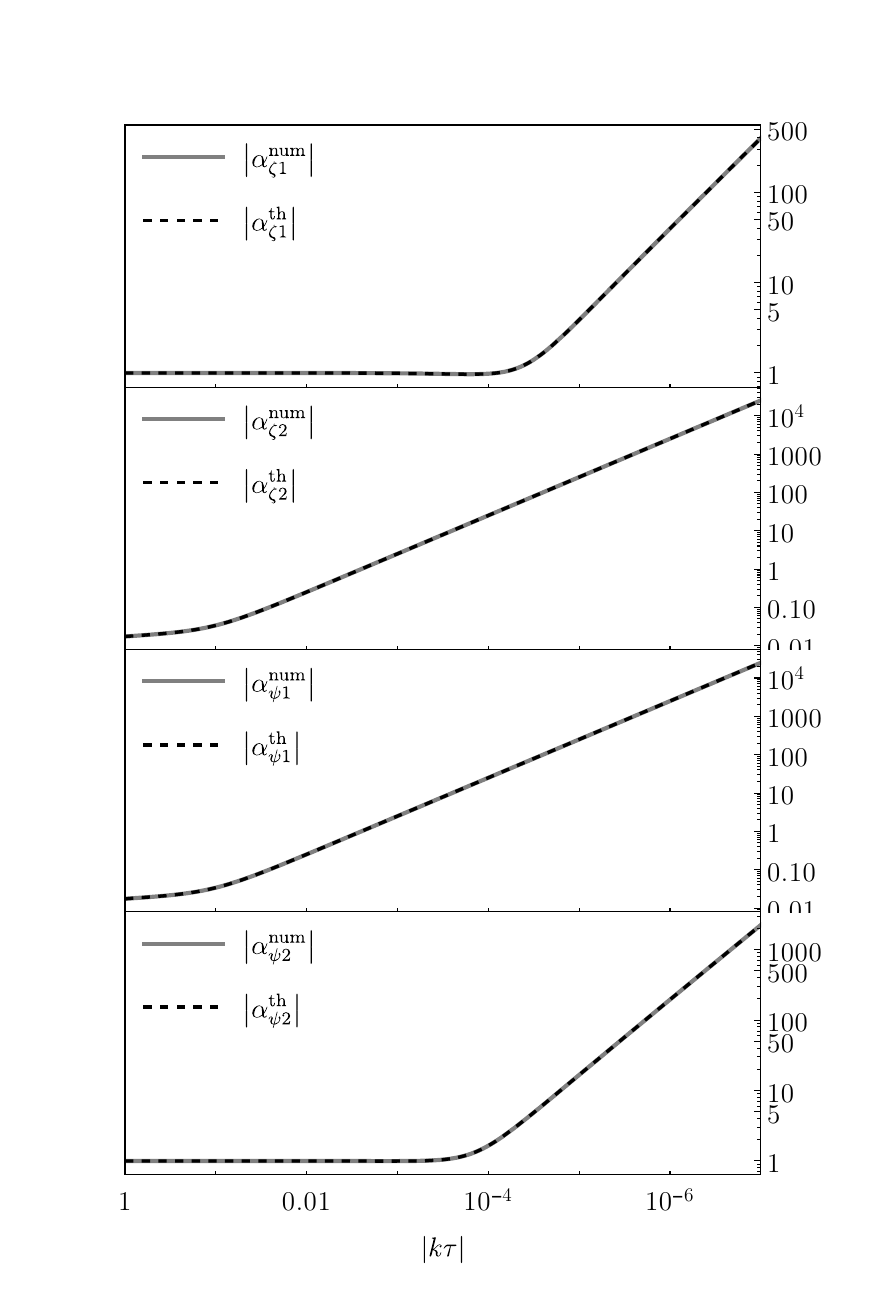}
    \caption{}
    \label{subfig:abs-a}
\end{subfigure}
\begin{subfigure}{0.49\textwidth}
    \hspace*{-1cm}
    \includegraphics[width=1.2\linewidth]{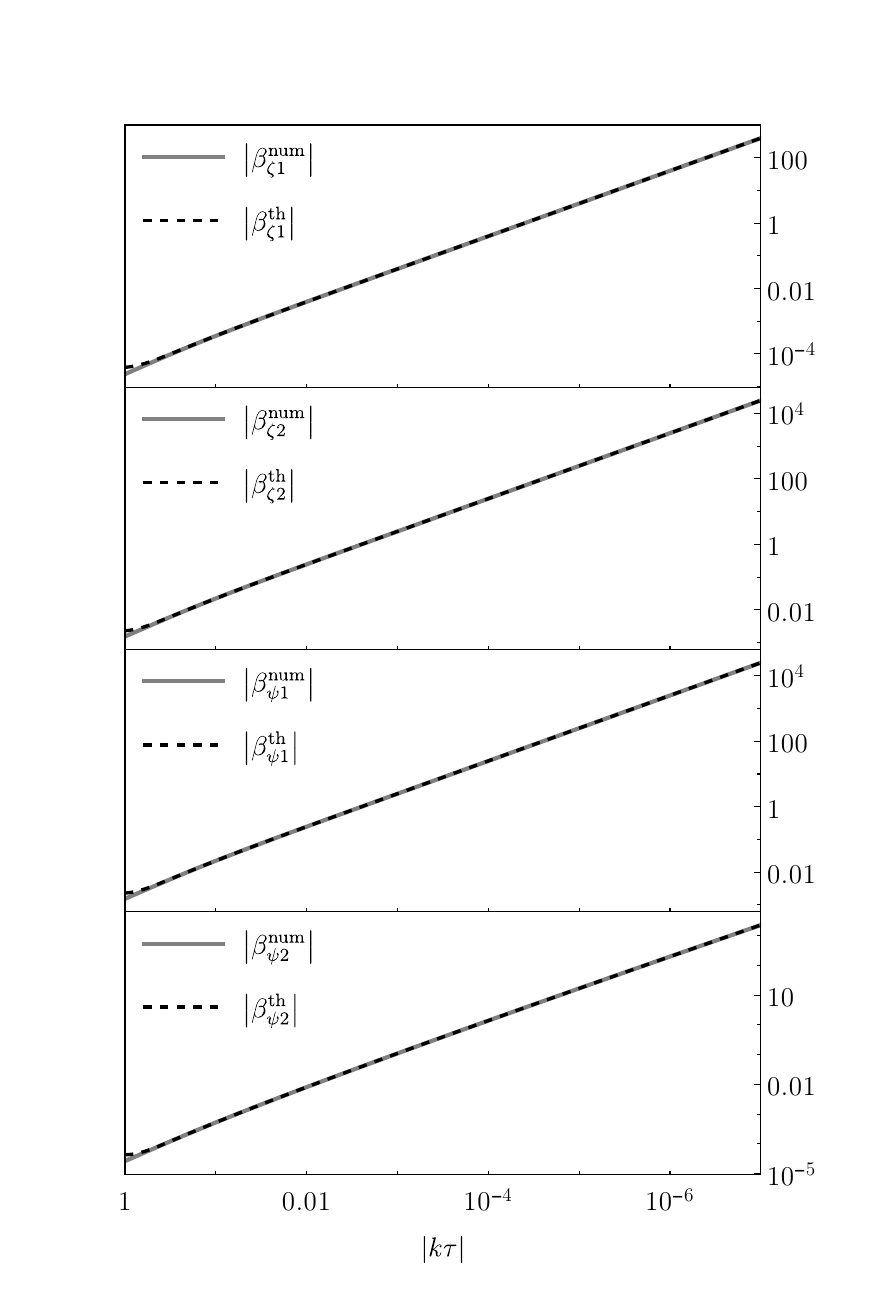}
    \caption{}
    \label{subfig:abs-b}
\end{subfigure}
\caption{ Comparison between the numerical solution of Eq.~(\ref{eq-alpha-beta-multi-1}) and the analytical expressions~(\ref{multibogoz-1}-\ref{multibogoz-2}) obtained by integrating Eq. (\ref{sol-Dyson-2}) to second order. Parameters set to $k|\tau_{\mathrm{ini}}|=10^{3}$, $\lambda=5\times 10^{-2}$.  }
\label{}
\end{figure}

\begin{figure}[h!]
\centering
\begin{subfigure}{\textwidth}
\hspace*{-1cm}
    \includegraphics[width=1.1\linewidth]{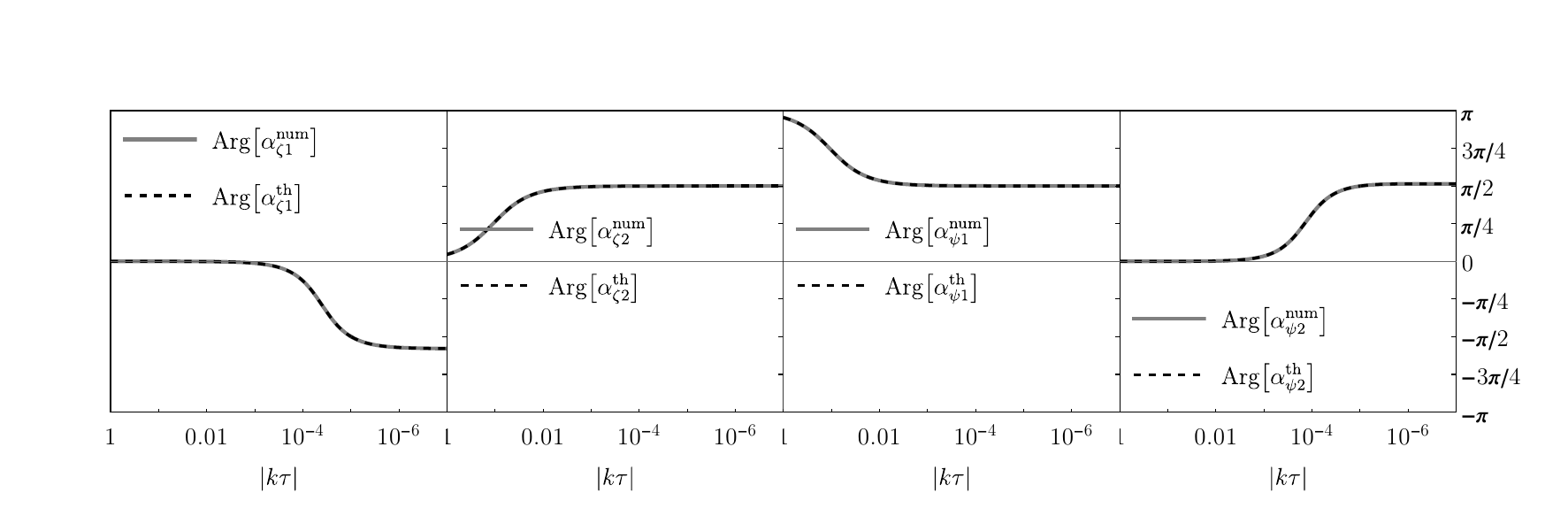}
    \caption{}
    \label{subfig:arg-a}
\end{subfigure}
\begin{subfigure}{\textwidth}
\hspace*{-1cm}
    \includegraphics[width=1.1\linewidth]{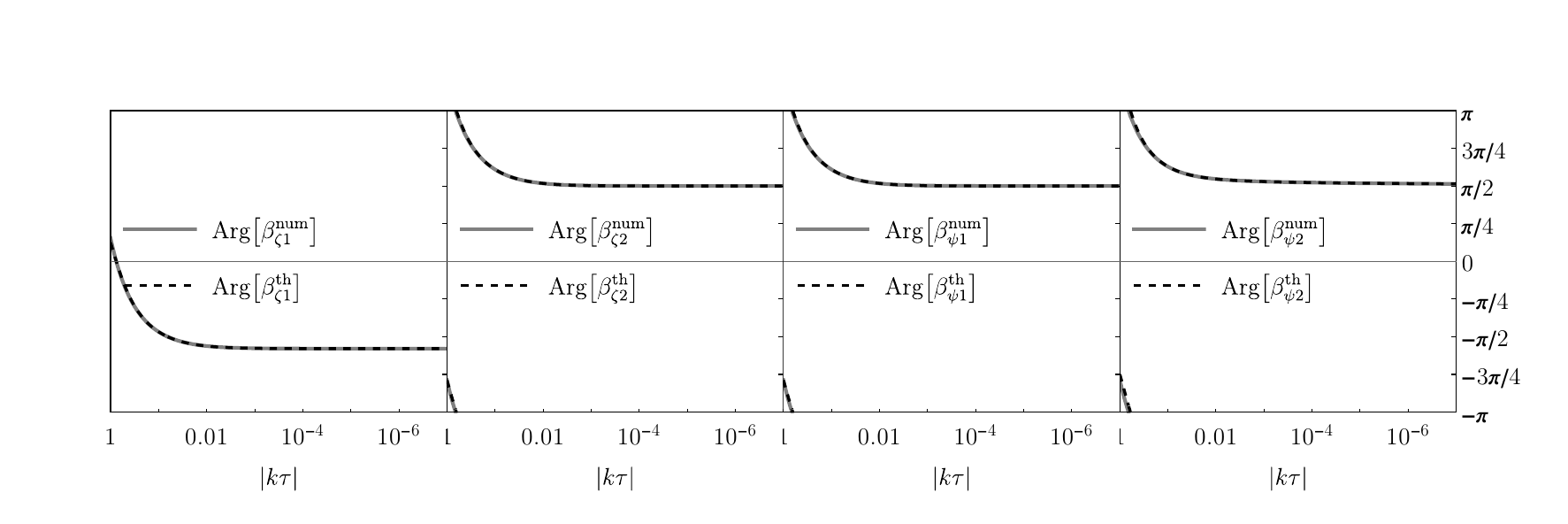}
    \caption{}
    \label{subfig:arg-b}
\end{subfigure}
\caption{ Comparison between the numerical solution of Eq.~(\ref{eq-alpha-beta-multi-1}) and the analytical expressions~(\ref{multibogoz-1}-\ref{multibogoz-2}) obtained by integrating Eq. (\ref{sol-Dyson-2}) to second order. Parameters set to $k|\tau_{\mathrm{ini}}|=10^{3}$, $\lambda=5\times 10^{-2}$.  }
\label{}
\end{figure}
In Figs.~1 and~2, we check these expressions against the numerical integration of Eq.~(\ref{eq-alpha-beta-multi-1}) finding excellent agreement.
It can be also checked that these expressions have the appropriate asymptotic form satisfied by Eqs.~(\ref{asympt-sol-alpha-beta-1})-(\ref{asympt-sol-alpha-beta-4}) derived in the previous discussion. This result implies that further corrections of higher order in $\lambda$ will not contain larger powers of $\ln (- k \tau)$. As a consequence, we can trust these solutions for $|\lambda| \ll 1$ even if $| \lambda \ln (- k \tau) | \gg 1$. 

We may now compute observables such as the power spectrum of $\zeta$ and $\psi$. Using Eqs.~(\ref{power-zeta-alpha}) and (\ref{power-psi-alpha}), we finally obtain
\bea
P_{\zeta} &=&  \frac{ H^2 }{ 4 \epsilon k^3} \Big[ 1 + \frac{\lambda^2}{12} (\pi^2-12) + \lambda^2 \big( w + \ln(-k\tau) \big)^2 \Big] , \label{power-zeta-alpha-2} \\
P_{\psi} &=&  \frac{ H^2 }{ 2 k^3} \Big[ 1 + \frac{\lambda^2}{6} (\pi^2 - 6)  \Big] ,  \label{power-psi-alpha-2} \\
P_{\zeta\psi} &=& \frac{-H^2}{2\sqrt{2\epsilon}k^3} \lambda \Big[ w + \ln(-k\tau) \Big]. \label{power-psi-zeta-alpha-2}
\eea
It can be appreciated that these observables are independent of $\tau_{\rm ini}$, as expected. Note that~\eqref{power-zeta-alpha-2} and \eqref{power-psi-alpha-2} do not fully coincide with those derived in Ref.~\cite{Achucarro:2016fby} (we have verified, however, that our expressions coincide with the numerical integration of Eq.~(\ref{eq-alpha-beta-multi-1}) with appropriate initial conditions).

\subsection{Particle spectrum in multifield models}
Having obtained the Bogoliubov coefficients that encode the dynamics of the two-field model, we are now in a position to compute the number-density of Hamiltonian eigenstates produced over the BD state due to the kinetic mixing induced by the non-geodesic motion. To do so we may substitute the $\beta_{Ia}$ coefficients found in~(\ref{multibogoz-1}-\ref{multibogoz-2}) into Eq.~\eqref{n_a}. 

As an example, let us write explicitly the short- and long-wavelength limits of the number density. In the $k\tau \gg 1$ regime, the integral simplifies to
\be
    n_{\bar a}(k,\tau) 
    =  
    n_\K^{\rm BD}(\tau) \left( 1 +  \frac{\lambda^2}{4}  \right) ,
\ee
with  $n_\K^{\rm BD}(\tau)$ the number-density of eigenstates of the BD state given in Eq.~\eqref{n-BD}.
In the long-wavelength limit, $\lambda < |k\tau| \ll 1$, we obtain
\be \label{n-abar-IR}
    n_{\bar a}(k) =n_\K^{\rm BD}(\tau) \bigg( 1 + \frac{\lambda^2}{8 \left( k\tau \right)^2} \Big[ 1 + \mathcal{O}\left( (k\tau)^2 \right) \Big] \bigg) .
\ee
The requirement $|k\tau|>\lambda $ ensures that the eigenfrequencies remain real, and thus we can interpret the result as particle production. 
Let us note that contrary to the Hubble-flow induced non-adiabaticity, the number density of particles in this case cannot be written as the superhorizon limit of a thermal distribution ---recall the comments around Eq.~\eqref{BE} and below Eq.~\eqref{P2P1}. This is consistent with the fact that $\lambda$ is a non-gravitational coupling.

Finally, we can also carry out the computation for a transient interaction. In such scenario, according to Eq. (\ref{eq-alpha-beta-multi-1}), Bogoliubov coefficients evolve until $\lambda$ is turned off, inducing excited states after the turn ends~\cite{Palma:2020ejf,Fumagalli:2020adf,Fumagalli:2020nvq,Fumagalli:2021mpc}. Since $\lambda=0$ by the end of inflation, we employ Eqs. (\ref{numb-density-dcpld-1},~\ref{numb-density-dcpld-2}) to calculate the eigenstate number-densities across all wavelengths. For $|k\tau|\ll 1$ (and $\mu=0$), we obtain
\begin{align}
    n_{\bar 1}(k) &= n_\K^{\rm BD}(\tau) \frac{P_\zeta}{P_{\zeta_0}} + \frac{1}{2} \frac{P_\zeta}{P_{\zeta_0}} - \frac{1}{2} , \\
    n_{\bar 2}(k) &= n_\K^{\rm BD}(\tau) \frac{P_\psi}{P_{\psi_0}} + \frac{1}{2} \frac{P_\psi}{P_{\psi_0}} - \frac{1}{2} ,
\end{align}
with the power spectra given by Eqs.~(\ref{power-zeta-alpha},~\ref{power-psi-alpha}) and $P_{\psi_0} = \epsilon  P_{\zeta_0} = H^2/ (4 k^3)$. We thus conclude that 
multifield interactions affect the power spectrum and the number-density in the same manner: since the piece proportional to $n^{\rm BD}$ dominates in the long-wavelength limit, we obtain $n_{\bar 1 , \bar 2 }/n^{\rm BD} \approx P_{\zeta,\psi}/ P_{\zeta_0,\psi_0} $. Notably, the strong amplification of the power spectrum observed in rapid-turn models~\cite{Palma:2020ejf,Fumagalli:2020adf,Braglia:2020eai} implies a corresponding enhancement in the particle content.

\section{Summary and concluding remarks}
\label{Sec:conclusions}
Particle production in curved spacetime is a ubiquitous process, albeit with ambiguous interpretations. Indeed, in order to study the phenomenon, one needs to make a choice: a vacuum state over which the excitations are considered, which, in turn, fixes a quantum Hamiltonian dictating their production rate. Each choice leads to different particle densities. 

Here, we have utilized an operator basis wherein the Hamiltonian is diagonal, allowing us to define particles as energy/momentum eigenstates. Following the system's canonical evolution via Bogoliubov transformations, we saw that while the adiabatic vacuum is indeed devoid of eigenstates at any initial time, the Bunch-Davies vacuum (which is devoid of particles only at the infinite past) leads to particle production encoding the spatial expansion of the Poincar\'e patch of de Sitter space. This is reflected in the fact that the eigenstate number-density of the BD state per physical momentum is constant ---recall Eq.~\eqref{N-phys}--- in direct analogy with the constant dimensionless power spectrum of a massless scalar. This is a consequence of the exact scale invariance of de Sitter space. 

In addition, the Bunch-Davies state can serve as a basis to expand any solution whose temporal dynamics deviates from exact (adiabatic) de Sitter. The coordinates in this basis are nothing but time dependent Bogoliubov coefficients that obey a first-order differential equation ---Eq.~\eqref{eq-alpha-beta-1}--- which can be solved perturbatively. A prime example of such a case in single-field inflation is the slow-roll scenario, where the non-adiabaticity is encoded in the non-vanishing slow-roll parameters $\epsilon$ and $\eta$. In this case, we saw that the number-density of energy/momentum eigenstates produced over the BD state, encodes a mild feature of the primordial curvature power spectrum: the observed spectral index, $n_s$. Another example worth studying in this category is the temporarily non-adiabatic dynamics taking place in the case of sharp features~\cite{Chluba:2015bqa}. Indeed, using the same methods, it should be possible to show that the number density in this case, encodes the featured power spectrum ---recall Eq.~\eqref{P2P1}. We are thus lead to conclude that the eigenstate content of the Bunch-Davies vector is a physical observable.   

Adopting this perspective, we went on to study energy/momentum eigenstates in the context of multifield inflation, specifically two-field models exhibiting non-geodesic motion. First, we saw that the multifield adiabatic basis leads to Hamiltonian eigenstates characterized by dispersion relations ---Eq.~\eqref{multifield-eigenfreq-c}--- identical to those found in the pertinent literature. These were well-known in the Minkowski (ultraviolet) limit of the model but using the Hamiltonian formalism allowed us  to rederive them in an exact manner. Given the dispersion relations, we then generalized the number-density of energy/momentum eigenstates in the multifield context ---Eq.~\eqref{n_a}. Particle production here is not due to quasi-de Sitter dynamics (even though this effect can also be taken into account ---see~\cite{Bianchi:2024jmn} for a single-field analogue) but instead results from the non-adiabaticity induced by the non-gravitational coupling of perturbations provided by the turning rate of the background trajectory. 

Having obtained the number density for any mode function, that is, of any Hilbert-space vector, one can ask the question of what is the particle content of the Bunch-Davies state. To do so we need to project the multifield solution onto the Bunch-Davies basis. This can be done in analogy with the single-field case, \emph{i.e.} using Boguliubov transformations adapted to the system at hand. 
The coordinates of this projection, that is, the set of Bogoliubov coefficients, obeys again a first-order differential equation ---Eq.~\eqref{eq-alpha-beta-multi-1}--- which constitutes a generalization of the single-field case. We derived perturbative analytical solutions of this equation ---Eqs.~(\ref{multibogoz-1}-\ref{multibogoz-2})--- and used them to compute the corresponding two-field eigenstate occupation numbers. Remarkably, the effects of the kinetic mixing are distinct from the gravitationally induced process (\emph{e.g.} particle production due to slow-roll) and can be isolated from such effects. This is evident from the different scaling of the number densities written in Eqs.~\ref{n-QDS} and \ref{n-abar-IR}: logarithmic vs power law. The logarithmic running arising from the $n_s\to1$ limit of Eq.~\ref{n-QDS} is characteristic of the soft breaking of scale invariance stemming from the quasi-de Sitter background. On the other hand, the non-gravitational coupling induces a hard scale dependence, which cannot be viewed as a deformation of the scaling dimension even in the weakly coupled, $\lambda<1$, regime. Note that this distinction on the scale dependence can serve as a tool for disentangling feature contributions in primordial spectra.

Our results can also have practical applications in any setup where the number-density of non-adiabatically produced particles, plays a central role. For example, one can compute the relic abundance of isocurvature excitations to judge if $\psi$ can constitute a dark-matter component~\cite{Kolb:2022eyn} (recall that $\psi$ is only massless during inflation in the class of models under consideration). Next, we can compute the energy density
$\rho_{\bar a} = a^{-4}\int \dd[3]k \, n_{\bar a} \omega_{\bar a}$ and see when this becomes larger than the energy density driving inflation, in which case backreaction on the geometry starts being non-negligible~\cite{Holman:2007na}. Given that we are in the weak-coupling regime $\lambda<1$, this quantity can place constraints on how long the non-geodesic motion can last before backreacting on the inflationary dynamics. On the other hand, the analytical expressions for the Bogoliubov coefficients can be used to compute the contribution of isocurvature particle production to the stochastic gravitational-wave background and the enhancement of non-Gaussian correlation functions~\cite{Holman:2007na,Meerburg:2009ys,Agarwal:2012mq,Aravind:2013lra,Parra:2024}. We leave these questions for future work.

\section*{Acknowledgments} 
GAP and NP acknowledge support from the Fondecyt Regular project number 1210876 (ANID). SS is supported by Thailand NSRF via PMU-B [grant number B37G660013].

\end{document}